\documentclass[preprint]{emulateapj}

\newcommand\etal{{\it et al. }}
\def\kms{\ifmmode{\rm km\,s^{-1}}\else\hbox{$\rm km\,s^{-1}$}\fi}
\def\ergs{\ifmmode{\rm ergs\,s^{-1}}\else\hbox{$\rm ergs\,s^{-1}$}\fi}
\def\farcs{\hbox{$.\!\!^{\prime\prime}$}}

\slugcomment{To be submitted to Astrophysical Journal}

\shorttitle{Dust and the Type II-P SN 2004et}
\shortauthors{Kotak et al.}

\begin{document}
\title{Dust and the Type II-Plateau supernova 2004et.}


\author{R. Kotak,\altaffilmark{1}
W. P. S. Meikle,\altaffilmark{2} 
D. Farrah,\altaffilmark{3}
C. L. Gerardy,\altaffilmark{4}
R. J. Foley,\altaffilmark{5}
S. D. Van Dyk,\altaffilmark{6}
C. Fransson\altaffilmark{7},
P. Lundqvist\altaffilmark{7},
J. Sollerman\altaffilmark{7,8}, 
R. Fesen\altaffilmark{9},
A. V. Filippenko\altaffilmark{5},
S. Mattila\altaffilmark{10},
J. M. Silverman,\altaffilmark{5}
A. C. Andersen,\altaffilmark{8}
P. A. H\"{o}flich\altaffilmark{4},
M. Pozzo\altaffilmark{11}, \&
J. C. Wheeler\altaffilmark{12}
}

\altaffiltext{1}{Astrophysics Research Centre, School of Mathematics 
and Physics, Queen's University Belfast, BT7 1NN, United Kingdom
r.kotak@qub.ac.uk}
\altaffiltext{2}{Astrophysics Group, Blackett Laboratory, Imperial
College London, Prince Consort Road, London SW7 2AZ, United Kingdom}
\altaffiltext{3}{Astronomy Centre, Dept. of Physics and Astronomy,
University of Sussex, Brighton BN1 9QH, United Kingdom}
\altaffiltext{4}{Department of Physics, Florida State University, 
315 Keen Building, Tallahassee, FL 32306-4350}
\altaffiltext{5}{Department of Astronomy, University of California, 
Berkeley, CA 94720-3411}
\altaffiltext{6}{Spitzer Science Center, 220-6 Caltech, Pasadena, CA 91125}
\altaffiltext{7}{Department of Astronomy, Stockholm University, 
AlbaNova, SE-10691 Stockholm, Sweden}
\altaffiltext{8}{Dark Cosmology Centre, Niels Bohr Institute, University 
of Copenhagen, Juliane Maries Vej 30, 2100 Copenhagen {\O}, Denmark}
\altaffiltext{9}{Department of Physics and Astronomy, 6127 Wilder Lab., 
Dartmouth College, Hanover, NH 03755}
\altaffiltext{10}{Tuorla Observatory, Department of Physics and Astronomy, 
University of Turku, V\"{a}is\"{a}l\"{a}ntie 20, FI-21500 Piikki\"{o}, 
Finland}
\altaffiltext{11}{Department of Earth Sciences, University College London, 
London WC1E 6BT, United Kingdom} 
\altaffiltext{12}{Astronomy Department, University of Texas at Austin, 
Austin, TX 78712}




\begin{abstract}
We present mid-infrared (MIR) observations of the Type~II-plateau
supernova (SN) 2004et, obtained with the {\it Spitzer Space Telescope}
between days~64 and 1406 past explosion. Late-time optical spectra are
also presented. For the period 300--795~days past explosion, we argue
that the spectral energy distribution of SN~2004et comprises (a) a hot
component due to emission from optically thick gas, as well as
free-bound radiation, (b) a warm component due to newly formed,
radioactively heated dust in the ejecta, and (c) a cold component due
to an IR echo from the interstellar-medium dust of the host galaxy,
NGC~6946. There may also have been a small contribution to the IR SED 
due to free-free emission from ionised gas in the ejecta.
We reveal the first-ever spectroscopic evidence for
silicate dust formed in the ejecta of a supernova.  This is supported 
by our detection of a large, but progressively declining, mass of SiO.
However, we conclude that the mass of directly detected ejecta dust
grew to no more than a few times $10^{-4}\,{\rm M}_{\odot}$.  We also
provide evidence that the ejecta dust formed in comoving clumps of
fixed size. We argue that, after about two years past explosion, the
appearance of wide, box-shaped optical line profiles was due to the
impact of the ejecta on the progenitor circumstellar medium and that
the subsequent formation of a cool, dense shell was responsible for a
later rise in the MIR flux.  This study demonstrates the rich,
multi-faceted ways in which a typical core-collapse supernova and its
progenitor can produce and/or interact with dust grains. The work 
presented here adds to the growing number of studies which do not 
support the contention that SNe are responsible for the large
mass of observed dust in high-redshift galaxies.
\end{abstract}


\section{Introduction}
\label{sec:intro}
\setcounter{footnote}{0}
\setcounter{section}{1}
\setcounter{subsection}{0}

Historically, the opening up of the electromagnetic spectrum beyond
the narrow visible region has always led to a deeper understanding of
physical phenomena.  Nowhere is this currently more evident than in
the mid-infrared (MIR) regime where the combination of vastly superior
sensitivity and spatial resolution afforded by the {\it Spitzer Space
Telescope\/} \citep{werner:04}, compared to previous infrared (IR)
satellites, has transformed the study of extragalactic point sources.

Ground-based MIR studies of supernovae (SNe) are extremely challenging --- 
if not impossible --- even from the highest-altitude facilities. This is 
due to a combination of conspiring factors: strong terrestrial
atmospheric absorption, generally high background in the MIR, and
the faintness of MIR emission from SNe at epochs of interest.

Currently, one of the most persistent questions relating to SN
research is whether core-collapse SNe could be major contributors to
the universal dust budget. Although this was suggested about 40 years
ago \citep{cernuschi:67,hw:70}, attempts to verify this hypothesis
observationally have, until recently, remained few and far between.
This has been partly due to a lack of sufficiently sensitive
instrumentation, and partly because, given the large amounts of dust
produced in the winds of low-mass stars in the local
Universe, the notion that SNe might produce significant dust was not
widely explored.  However, evidence in favor of enormous amounts of
dust ($\gtrsim 10^{8}\,{\rm M}_\odot$) in galaxies at high redshifts 
($z\gtrsim 6$) is mounting, and comes from a variety of observations
such as sub-mm observations of the most distant quasars \citep{bertoldi:03},
obscuration by dust of quasars in damped Ly-$\alpha$ systems (DLAs)
\citep{pei:91}, and measurements of metal abundances in DLAs
\citep{pettini:97}.  The existence of large amounts of dust when the
Universe was relatively youthful poses problems for a low-mass star
origin of dust, since the main-sequence evolution timescales of these
stars (up to 1\,Gyr) begin to become comparable to the age of the
Universe.  SNe arising from short-lived population III stars could be
a viable alternative.  Although estimates of the amount of dust
produced per supernova are sensitive to the choice of initial mass
function and grain destruction efficiencies, the current consensus is
that each SN need only produce 0.1--1\,${\rm M}_\odot$ of dust to 
account for the high-redshift observations \citep{dgj:07,meikle:07}.
However, for the handful of recent SNe for which dust-mass
estimates exist, these invariably fall short of the amounts cited
above by 2--3 orders of magnitude.

There are at least two direct methods for inferring the presence of dust 
in the ejecta of recent SNe. The first relies on the attenuation of the
red wings of optical or near-infrared (NIR) ejecta lines during the
nebular phase, betraying the presence of new dust that has
condensed in the ejecta. The signature is usually pronounced, and has
not yet been attributed to any other effect.  While this method has
been used to infer the presence of freshly formed dust, quantitative
information on the amount, composition, and geometry has remained
elusive. The other method is to observe the thermal emission
from dust grains.  Although over a dozen SNe have exhibited
NIR excesses, the challenge so far has been to distinguish between
pre-existing circumstellar dust heated by the SN, and new dust
condensing in the ejecta. The former scenario (i.e., the IR echo)
predicts a distinctive flat-topped light-curve shape
\citep[e.g.,][]{be:80,dwek:83}, thus providing one means by which to
differentiate between the two cases.  The ideal observational
requirement for this method to work is that of a long time-series of
MIR data. Additionally, multi-wavelength MIR data provide information
on the evolution of the spectral energy distribution, thereby
constraining the dust temperature, emissivity, and geometry.

In the case of SN~1987A, both techniques were employed.  However, 
even the highest value obtained was only $7.5 \times 10^{-4}\,M_\odot$
\citep{ercolano:07}. \citet{pozzo:04} used the attenuation method to
infer a dust mass exceeding $2\times10^{-3}$~{\rm M}$_{\odot}$ in the
Type~IIn SN~1998S. However, such events are relatively rare.  For the
Type II-plateau (II-P) supernova SN~1999em, \citet{elmhamdi:03} used the 
attenuation method to obtain a lower limit to the dust mass of just
$\sim10^{-4}\,{\rm M}_\odot$.  More recently, using {\it Spitzer}
observations, the thermal-emission approach has been applied to the
Type II SN~2003gd.  \citet{meikle:07} found that the MIR flux at
about 16~months was consistent with emission from just
$4\times10^{-5}\,{\rm M}_\odot$ of newly condensed dust in the ejecta.  
They also showed that the claim by \citet{sugerman:06} of a much larger
mass of new dust at a later epoch was unsupported by the data.

Evidence for dust condensation in SN ejecta may also be acquired via
the study of supernova remnants (SNRs). However, MIR and far-IR (FIR)
studies of SNRs have tended to find low dust masses.  For example, for
the Cassiopeia~A SNR, \citet{hines:04} used {\it Spitzer} images at
24~$\mu$m and 70~$\mu$m together with other data to deduce a dust mass
of $3\times10^{-3}\,{\rm M}_\odot$ at around 80~K in Cas~A. From {\it
Spitzer} spectroscopy, \citet{rho:08} deduce a factor of 10 larger
dust mass, and suggest that the difference is due to their use of
additional grain compositions necessary to fit the spectra.  Three
studies based on {\it Spitzer} data of the young SNR 1E 0102.2--7219
reveal dust masses that vary by roughly two orders of magnitude:
\citet{stanimirovic:05} find no more than $8\times10^{-4}\,{\rm
M}_\odot$ of dust at $\sim$120~K associated with the remnant, while
\citet{sandstrom:09} and \citet{rho:09} derive dust masses of
$3\times10^{-3}\,{\rm M}_\odot$ at about 70~K, and $10^{-2}\,{\rm
M}_\odot$ of dust at $\sim$60~K, respectively.
Estimates for the Crab and Kepler's SNRs based on {\it Spitzer} 
MIR data are similar, with $10^{-3}-10^{-2}$~{\rm M}$_\odot$ of dust
for the former by \citet{tem:06}, and $3\times10^{-3}\,{\rm M}_\odot$ 
at $\sim$85~K for the latter by \citet{blair:07}.
%
Thus, the bulk of MIR/FIR studies of SNRs yield dust
masses which are at least a factor of 10 less than the 
{\it minimum} required to account for the high-redshift dust.

It is possible that most of the SNR dust is very cold ($\lesssim$30~K)
and hence has escaped detection in these studies.  Attempts have been
made to detect very cold dust using sub-mm measurements of the Cas~A
and Kepler SNRs.  Observations by \citet{dun:03} using {\it SCUBA} led
them to claim that at least 2~M$_\odot$ of dust formed in the
supernova, with their model including a dust component at 18~K.
However, \citet{kra:04} used the same data together with observations
from {\it Spitzer} to show that most of the emission originates from a
line-of-sight molecular cloud, and not from dust formed in
Cas~A. Nevertheless, in a recent study of 850~$\mu$m polarization in
Cas~A, \citet{dunne:09} still estimate a dust mass of
$\sim$1M$_\odot$. However, this estimate is subject to significant
uncertainties in 850~$\mu$m emissivity as well as in grain alignment
and morphology. In particular, they point out that iron
\citep{dwek:04} or graphite needles could reduce the dust mass
required to explain the sub-mm observations. For the Kepler SNR,
\citet{morgan:03} used {\it SCUBA} to infer a dust mass of
$\sim1\,{\rm M}_\odot$, where the model fit included a dust component
at 17~K. However, they subsequently revised this dust mass downward by
a factor of two in the sub-mm study of \citet{gomez:09}.  We conclude
that, so far, the direct sub-mm evidence in favor of large masses of
ejecta dust is relatively weak.  On the other hand, several studies
have highlighted the destruction of dust grains by the reverse shock
as the supernova ejecta crashes into the ambient medium and makes
the transition into the remnant phase
\citep[e.g.][]{williams:06,nozawa:07}.  This implies that the dust
masses that condense in SN ejecta at early epochs (few years) should
be several times higher than during the remnant phase.

All of the above findings have rekindled the debate regarding the role
of SNe as dust producers, and vigorous efforts are underway to address
this question.

{\em Spitzer's \/} coverage of the likely peak of thermal emission due
to dust, together with the availability of flexible scheduling of
observations, provided an opportunity to test the SN ejecta
dust-condensation hypothesis, at least for local supernovae.  We have
therefore been pursuing a campaign of MIR observations for a 
sample of core-collapse SNe within the framework of the
Mid-Infrared Supernova Consortium (MISC).

Fortuitously, there have been several nearby supernovae since
the launch of {\em Spitzer \/} in 2003, which have also been
well monitored at other wavelengths.  In \citet{kotak:05} we reported
early results on SN~2004dj, the first MIR observations of SN ejecta
since those of SN~1987A.  {\em Spitzer \/} studies by \citet{barlow:05}
and \citet{meikle:06} of the Type~II-P SN~2002hh revealed the
complicated nature of its surroundings.  However, both sets of authors
concluded that the MIR flux from this SN was dominated by emission
from a dusty circumstellar medium (CSM) or nearby molecular cloud, 
driven by the SN luminosity. This radiation swamped any prospects of 
detecting newly formed ejecta dust in this case.

In what follows we present the most extensive MIR dataset of any SN to
date, surpassed only in the timespan of observations by the
Type~II-pec SN~1987A, which has been observed at MIR wavelengths as
late as 18 years after explosion \citep{bouchet:06}.  Moreover, it can
be argued that since SN~2004et is of Type~II-P, the most
common of all core-collapse SN types, the MIR work presented here is
of particular relevance to the general understanding of the role
played by SNe in the formation of dust grains.

\begin{figure*}[!t]
\begin{centering}
\includegraphics[height=0.44\textwidth,clip=]{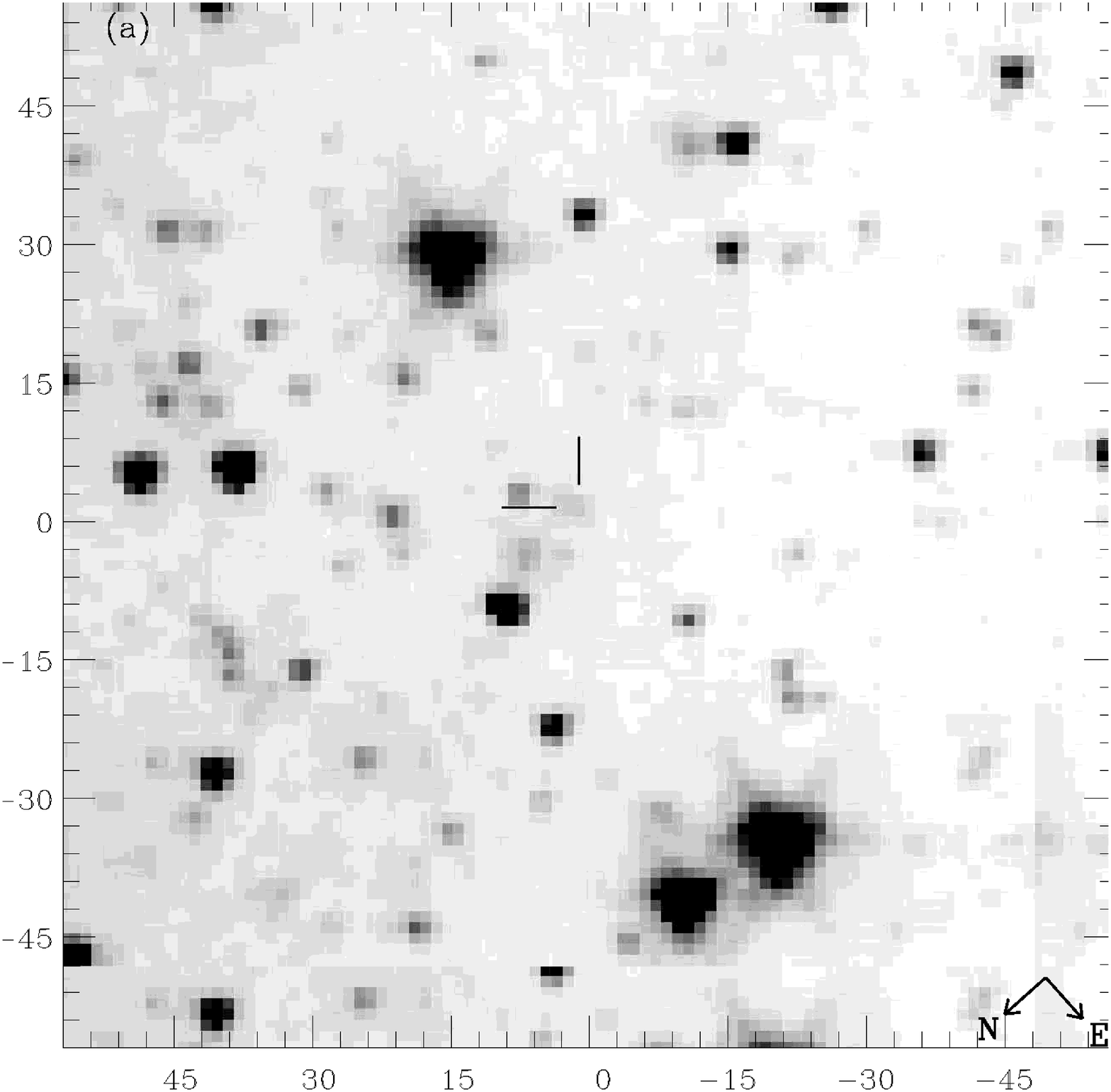}
\includegraphics[height=0.44\textwidth,clip=]{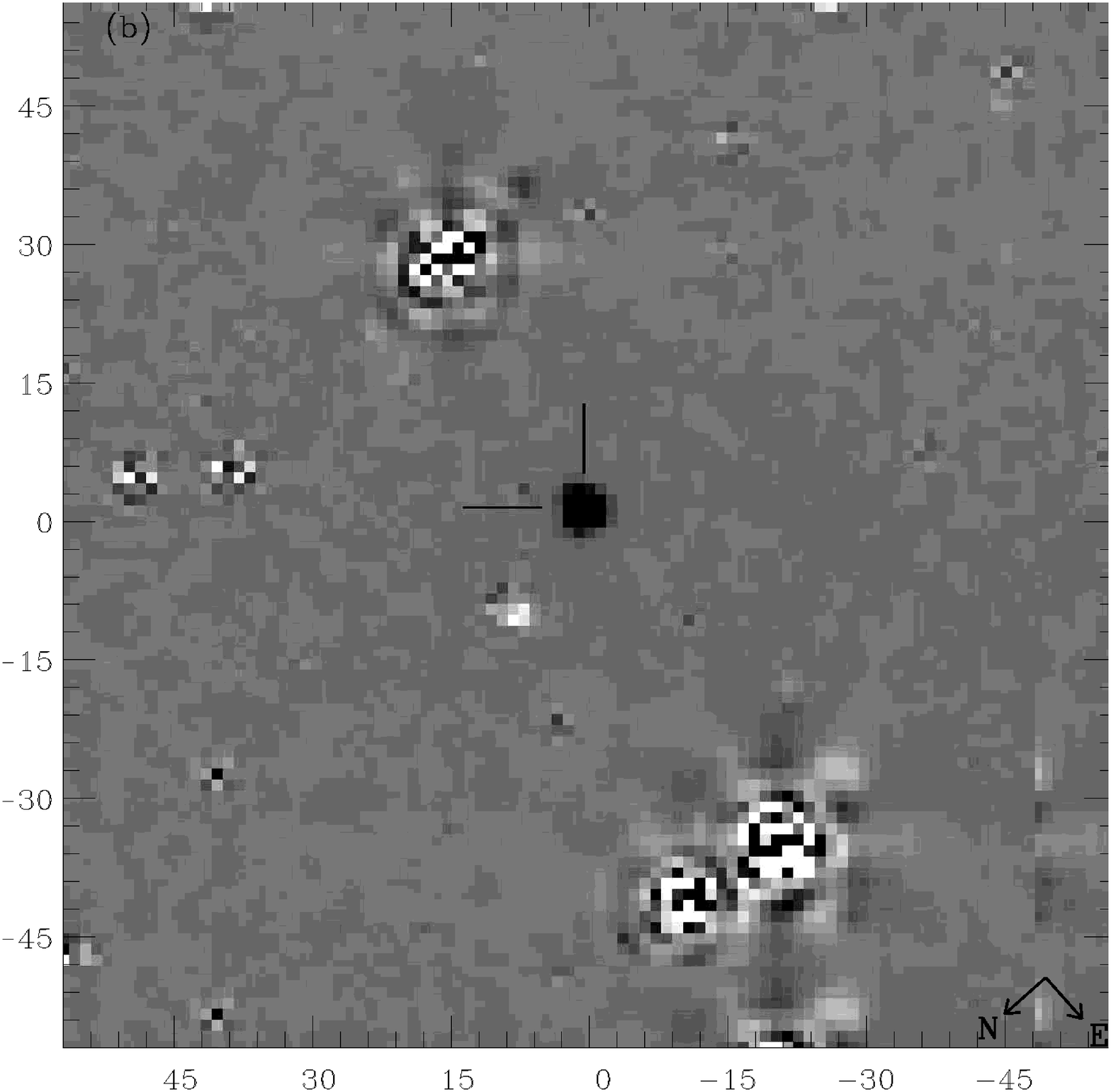}
\includegraphics[height=0.44\textwidth,clip=]{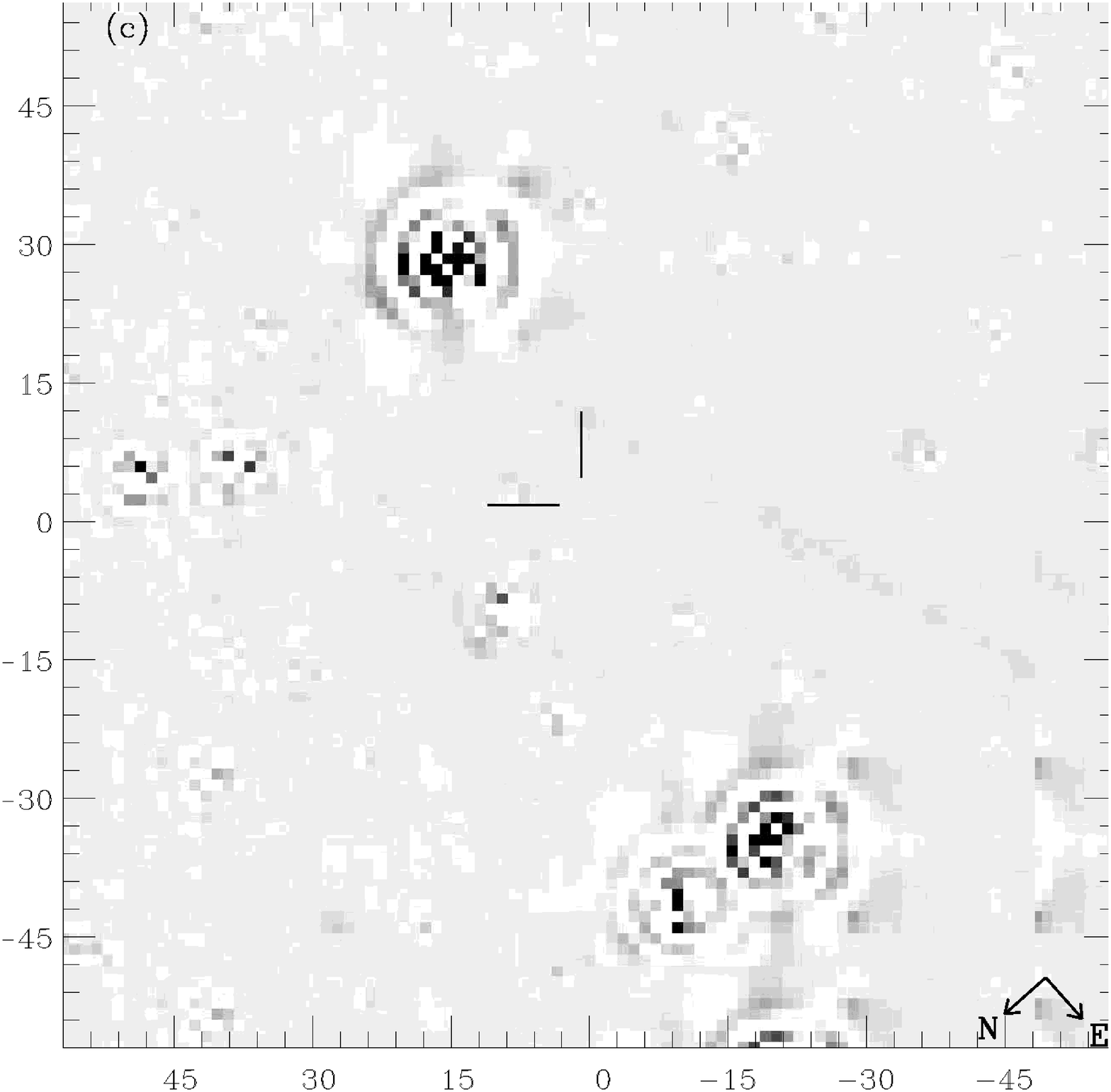}
\includegraphics[height=0.44\textwidth,clip=]{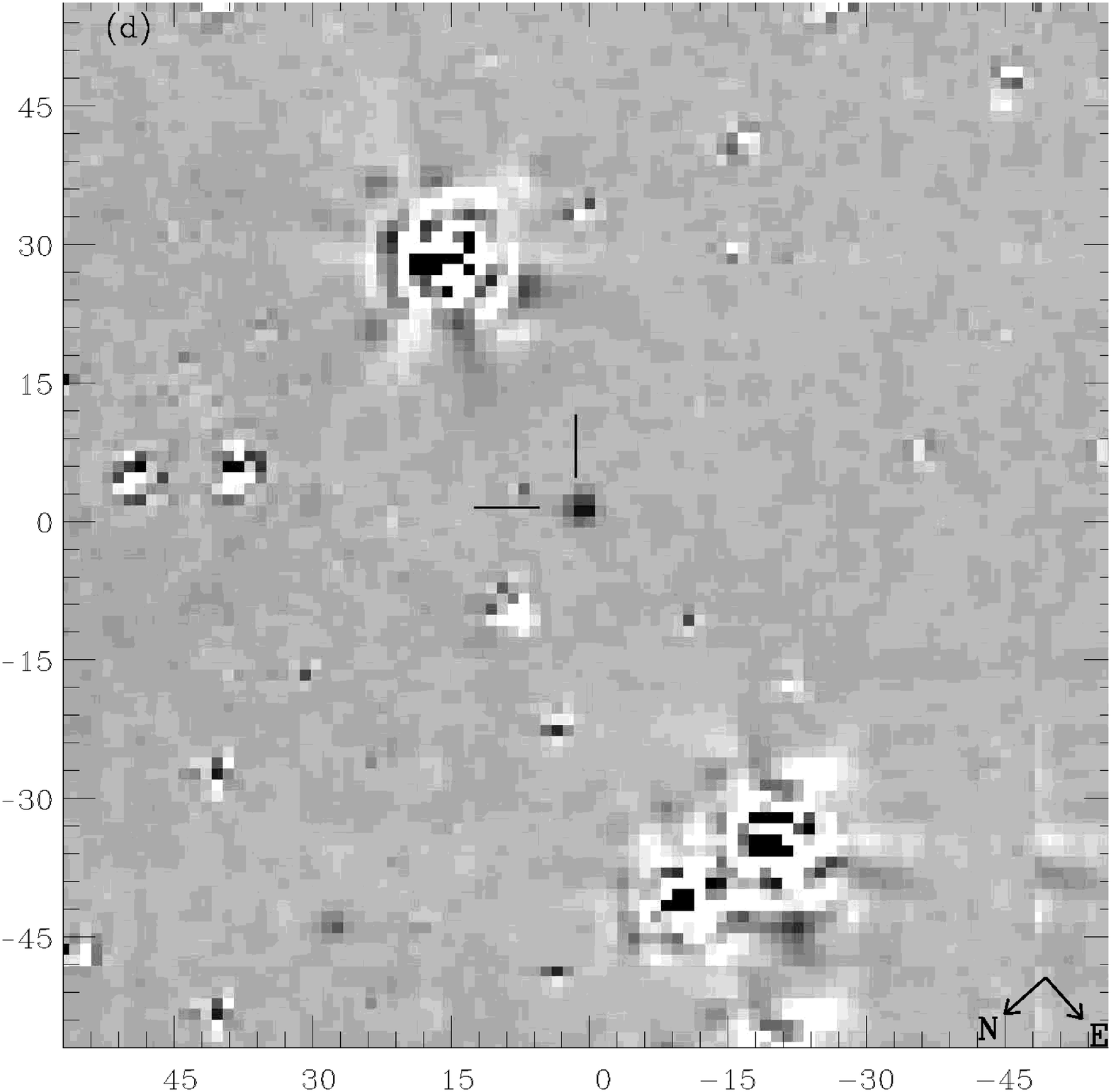}
\caption{Subsections of the immediate field around SN~2004et 
         at 3.6\,$\mu$m. Both axes are given in arcsec, and the
         field is only approximately centred on the supernova.
         Panel (a) shows the pre-explosion image, while panels 
         (b), (c), and (d) show template-subtracted frames of 
         the same field at 300, 795, and 1222\,days, respectively, 
         revealing the dramatic changes in
         brightness over this timespan (see also Table \ref{tab:phot}).
         Short dashes mark the position of the SN
         ($\alpha_{\mathrm{J2000}}=20^{h}35^{m}25^{s}.33;\,\,
         \delta_{\mathrm{J2000}}= +60\degr 07'17''.7$), which lies
         247$''.1$\,E and 115$''.4$\,S of the galaxy nucleus.
         \label{fig1}}
\end{centering}
\end{figure*}

\subsection{SN~2004et}
\label{sec:04et}

SN~2004et was discovered on 2004 Sep. 22.98 (UT dates are used
throughout this paper) by S. Moretti
\citep{zwitter:04} and is the eighth supernova to occur in NGC~6946 
since 1917. Following \citet{li:05}, we adopt an explosion date
(epoch 0~days) of 2004 Sep. 22.0 (MJD = JD-2400000.5 = 53270.0). 
The distance to NGC~6946 has been estimated using several different
techniques. However, as pointed out by \citet{li:05}, the uncertainty
is large. For consistency with our earlier work on SN~2002hh in the
same host galaxy \citep{meikle:06} a distance of 5.9~Mpc is adopted
\citep{karachentsev:00}. Soon after discovery, the SN was spectrally
classified as a young Type~II-P SN, exhibiting characteristic broad
P~Cygni profiles in the Balmer lines \citep{zmm:04,filippenko:04}.
Photometric and spectroscopic monitoring programs began in earnest
soon thereafter. Results from optical monitoring are summarized by
\citet{sahu:06} and \citet{misra:07}. 

Based on pre-explosion imaging,
\citet{li:05} suggested a yellow supergiant as a candidate progenitor
for SN~2004et. This has since been challenged by \citet{crockett:09},
who show that the point source seen in ground-based images comprises
at least three separate sources. They argue that the true progenitor
is actually an M-type red supergiant.
X-ray \citep{rho:07} and radio \citep{martividal:07} observations at 
early times have revealed the presence of a progenitor wind rendered 
detectable through the impact of the fastest-moving ejecta.

\section{Mid-Infrared Observations of SN~2004et}
\label{sec:obs}

SN 2004et was observed using the full suite of {\em Spitzer \/}
instrumentation. Images at 3.6, 4.5, 5.8, and 8.0\,$\mu$m were
obtained with the Infrared Array Camera (IRAC), at 16 and 22\,$\mu$m
with the Infrared Spectrograph (IRS) Peak-up Array (PUI), and at
24\,$\mu$m with the Multiband Imaging Photometer for Spitzer
(MIPS). Spectra between 5.2 and 37~$\mu$m were acquired with the
IRS in low-resolution mode. Imaging observations spanned epochs 64.7
to 1406.0~days. These observations were drawn from a number of
programs in addition to our own, and are catalogued in Table
\ref{tab:phot}. Several pre- and post-explosion 3.6~$\mu$m
IRAC images are shown in Fig. \ref{fig1}.

\subsection{Mid-Infrared Photometry of SN~2004et}
\label{sec:phot}

We used the post-basic calibrated data (PBCD) products provided 
by the {\em Spitzer \/} pipeline.  
The serendipitous pre-explosion IRAC and MIPS images from the SINGS
program showed evidence for spatially extended emission at and near
the location of the supernova.  We therefore subtracted these
``templates'' from all epochs of post-explosion imaging before
proceeding with our flux measurements.  The image matching and
subtraction was performed as implemented in the ISIS v2.2 image
subtraction package \citep{alard:98,alard:00}, and modified in a
manner analogous to that described by \citet{meikle:06}.

\begin{figure}[!t]
\begin{centering}
\includegraphics[height=0.5\textwidth,clip=]{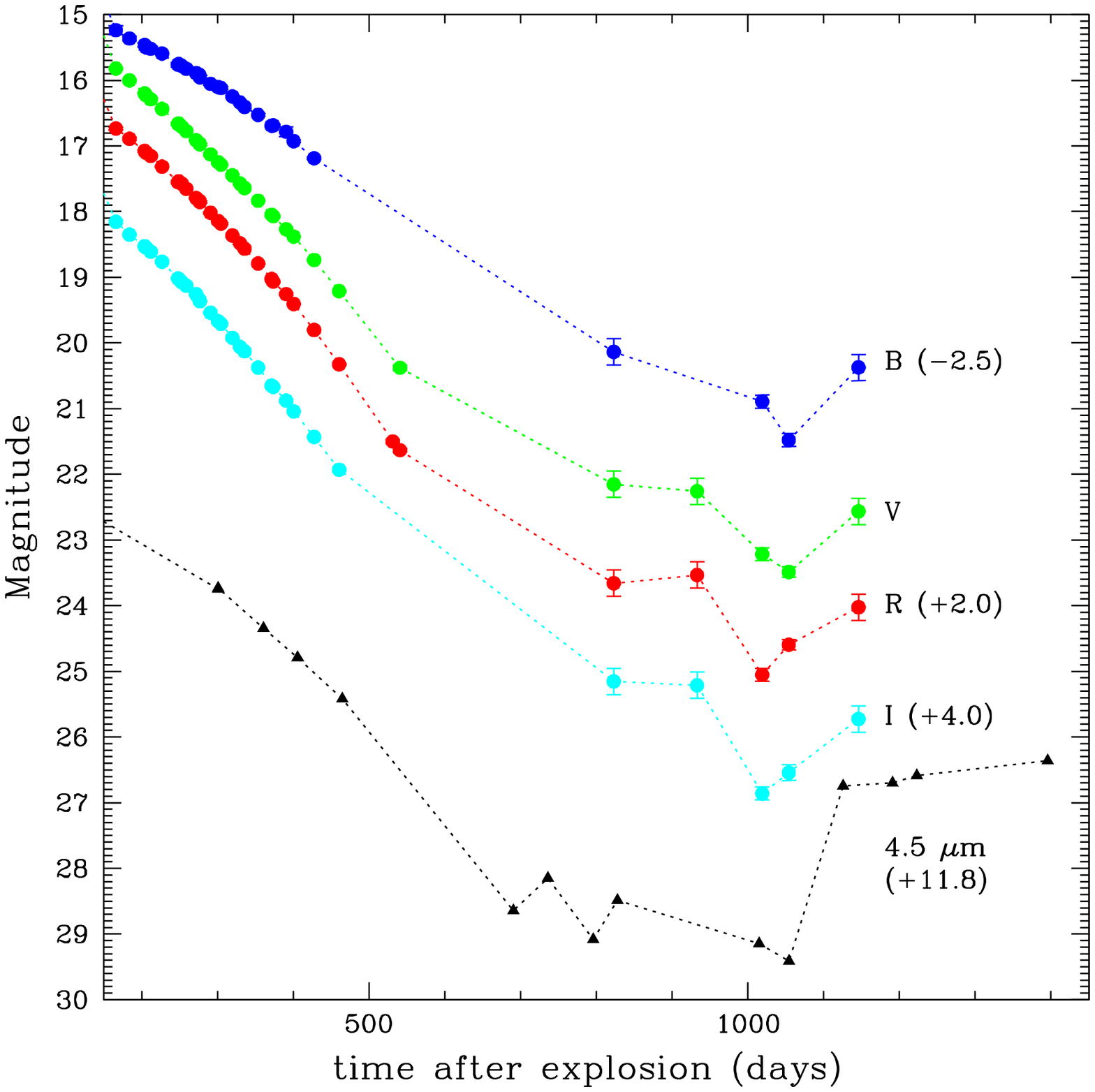}
\caption[]{Mid-infrared template-subtracted light curves of
SN~2004et. The upper limits are at $2\sigma$. For clarity the light
curves have been shifted vertically by the factors shown in brackets.
\label{fig2}}
\end{centering}
\end{figure}

Aperture photometry was performed on the background-subtracted IRAC
and MIPS images using the Starlink package GAIA \citep{draper:02}.  A
circular aperture of radius 5$''$ was used for the photometry. This
was chosen as a compromise between minimizing the effects of the
residual ridge emission at the SN location at late epochs in the IRAC
and MIPS data, and minimizing the the size of aperture correction
needed in the final flux determination.  The aperture radius
corresponds to a distance of $\sim$145~pc at SN~2004et.  Residual
background in the template-subtracted IRAC and MIPS images was
measured and subtracted by using a clipped mean sky estimator, and a
concentric sky annulus having inner and outer radii of 1.5 and 2 times
the aperture radius, respectively.  Aperture corrections were derived
from the IRAC and MIPS point-response function frames available from
the {\it Spitzer Science Center\/}, and ranged from a factor of 1.08 
at 3.6\,$\mu$m to a factor of 2.12 at 24\,$\mu$m. Fluxing errors due 
to uncertainties in the aperture corrections are about $\pm5\%$.  At
earlier epochs the aperture was centered by centroiding on the SN
image. At later epochs, when the supernova was faint, the aperture was
centered using the WCS coordinates, with the SN position given by the
radio observation of \citet{stockdale:04}.

For the PUI images (16 and 22\,$\mu$m), no pre-explosion data were
available. We carried out standard aperture photometry on the PBCD
images using the same sizes of aperture and annulus as for the IRAC
and MIPS measurements.  The residual background was then estimated as
follows.  Using the same aperture and sky annulus, we measured the net
fluxes at the SN location in the pre-explosion IRAC and MIPS images
(see Table~\ref{tab:phot}).  The residual background at 16 and
22\,$\mu$m was then found by blackbody interpolation between the IRAC
and MIPS wavelengths and subtracted from the previously measured PUI
fluxes. The resulting SN photometric measurements are listed in Table
\ref{tab:phot} and displayed as light curves in Fig. \ref{fig2}. The
uncertainties shown are statistical only.

\subsection{Mid-Infrared Spectroscopy of SN~2004et}
\label{sec:spec}

Low-resolution ($R \approx 60$--127) MIR spectroscopy of SN~2004et
covered seven epochs between 294 and 1385~days, and was obtained using
the Short-Low (SL; 5.2--14.5\,$\mu$m) and Long-Low (LL; 14--38\,$\mu$m)
modules of the Infrared Spectrograph \citep[IRS][]{houck:04}. The log
of observations is given in Table~\ref{tab:speclog}.

We began our reduction of the SL and LL spectra with the basic
calibrated data (BCD) frames. Starting with these frames, we first
subtracted the sky background by differencing between the two nod
positions. All subsequent steps (extraction, wavelength calibration, 
and flux calibration) were carried out using the optimal
extraction mode of the {\em Spitzer\/} Custom Extractor tool, SPICE.  
Given the faintness of the
target by day~1212, we also experimented with subtracting the sky
background by differencing between the adjacent order and carrying out
subsequent extraction steps using the Spectroscopic Modeling,
Analysis, and Reduction Tool (SMART). We found the two extractions to
be consistent but with the latter producing a cleaner spectrum, so we
used this extraction. A possible reason for this difference may be
that order-order sky subtraction is more suited for removing background
emission when the background is high and variable; see below.

Despite this careful reduction procedure, the fluxes of the IRS
spectra and the IRAC photometry were not completely consistent.  The
main problem was that the SN lay on a ridge of emission in the host
galaxy (Fig. \ref{fig1}a). While this could be removed for the IRAC
and MIPS images using the template method described above, such a
technique was not possible for the spectra. Furthermore,
this problem was aggravated by differences in the fixed sizes of the
spectrograph aperture slits and the circular apertures used for the
image photometry. An additional, though smaller, problem was that the
spectra were generally taken a few days before or after the images, 
during which time the SN flux changed. 

In order to correct for these difficulties, we calibrated the IRS
spectra against nearly contemporary photometry in the 8~$\mu$m
band, chosen since it was completely spanned by the short-low (SL)
spectrum.  The IRAC 8~$\mu$m transmission function was multiplied by
the MIR spectra and by a model spectrum of Vega. The resulting MIR
spectra for the SN and for Vega were integrated over wavelength and
compared. Hence, the total SN spectral fluxes in the 8~$\mu$m band
were derived. These were then compared with the temporally nearest
8~$\mu$m photometry to derive scaling factors by which the spectra
were multiplied.  Correction factors were, respectively, $\times0.82$,
$\times0.84$, $\times0.75$, $\times0.76$, $\times0.37$, $\times1.0$
and $\times0.94$ for SL spectra at 294, 349, 450, 481, 823, 1212 and
1385~days, respectively. The IRAC photometry epochs to which these
spectra were scaled were 300, 360, 464, 464, 795, 1222 and 1395~days,
respectively.

The bulk of the corrections was due to errors in the original spectral
flux calibration.  Apart from the 823~day spectrum\footnote{The most
contemporary set of IRAC data to the day~823 spectrum was actually
obtained only five days later on day~828. However, owing to the
weakness of the day~828 $8~\mu$m flux and the consequent difficulty of
reliably correcting for the strong, complex background even after
template subtraction, we decided to scale the day~823 spectrum to
day~795 when the SN was considerably brighter.}, only a small part of
these corrections was due to the shift from the spectral epochs to the
contemporary photometry epochs.  The large correction at day~823 may
have been due to the faint SN flux around this time and the
consequently strong effect of the residual ridge emission.  We did not
detect the source in SL order 2 on day~1212, so the spectrum only
covered about 2/3 of the 8~$\mu$m IRAC band. Within this restriction,
we found that the spectrum provided a match to the 1222~day photometry
without any scaling.

\begin{figure}[!t]
\begin{centering}
\includegraphics[height=0.5\textwidth,clip=]{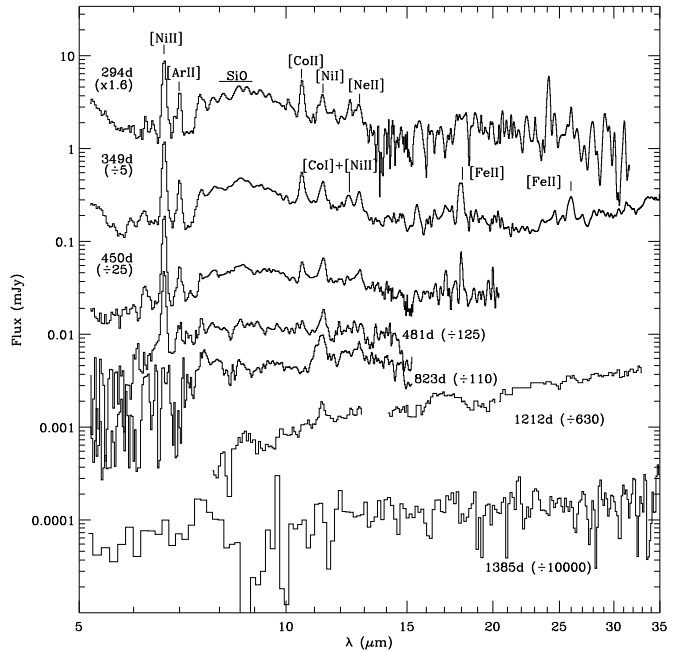}
\caption{MIR spectra of SN~2004et. While the epochs shown in
the diagram correspond to the original spectroscopic ones, the
respective spectra have been scaled to match contemporary IRAC
epochs at (respectively) 300, 360, 464, 464, 795, 1222 and 1395~days
(see \S~2.2).  For clarity the spectra have been shifted
vertically by the amounts shown in brackets.
\label{fig3}}
\end{centering}
\end{figure}

For five of the spectral epochs (294, 349, 450, 1212 and 1385~days),
usable contemporary long-low (LL) spectra, typically spanning
$\sim$14--30~$\mu$m, were also available. The first three had a small
overlap with the SL spectra in the 14~$\mu$m region.  These were
scaled to bring them into agreement with the SL spectra in this
region.  The LL correction factors were, respectively, $\times0.7$,
$\times0.5$, and $\times0.45$.  No SL-LL overlap was available for
spectral epochs 1212 and 1385~days.  Instead, theses spectra were
compared with contemporary 24~$\mu$m fluxes obtained with MIPS.  For
day~1212 the 14--21.3~$\mu$m portion required no correction to provide a
match to the photometry.  The 19.6--35~$\mu$m portion was scaled by a
factor of 0.75 to match the overlap with the shorter wavelength
section. For day~1385 the LL portion was scaled by $\times0.63$ to match
the contemporary photometry.  The spectra are plotted in
Fig. \ref{fig3}.  Note that, while the epoch labels shown in
Fig. \ref{fig3} correspond to the original spectroscopic ones, the
respective spectra have been scaled, via the above procedure, to match
contemporary photometry.  These corrections are important for the
analysis below which makes use of both the spectra and the photometry.
The day~481 spectrum was not used in the analysis below. This was
because of (a) the lack of a useable LL component, and (b) the close
temporal proximity of the day~481 SL spectrum to that on day~450.

\subsection{Very Late-Time Optical Spectroscopy of SN~2004et}
\label{sec:optspec}

Optical spectra ($\sim$4000 -- 9000~\AA) were taken using the 
Low-Resolution Imaging Spectrometer \citep[LRIS;][]{oke:95} mounted 
on the Keck I 10-m telescope on 2006 Dec. 25 and 2007 Nov. 12 (days~823 
and 1146, respectively) and the Deep Imaging Multi-Object Spectrograph 
\citep[DEIMOS;][]{faber:03} mounted on the Keck II 10-m telescope on 2007 
Apr. 14 (day~933).
We used a $1\farcs0$ wide slit which was aligned
along the parallactic angle to reduce differential light losses
\citep{filippenko:82}.  All spectra were reduced using standard
techniques with CCD processing and optimal extraction for the LRIS
data using IRAF.
We obtained the wavelength scale from low-order polynomial fits to
calibration-lamp spectra.  Small wavelength shifts were then applied
to the data after cross-correlating a template sky spectrum to the 
night-sky lines that were extracted with the SN. Using our own IDL routines, 
we fit spectrophotometric standard-star spectra to the data in order 
to flux calibrate our spectra and remove telluric lines 
\citep{wh:88}.

\begin{figure}[!t]
\begin{centering}
\includegraphics[height=0.48\textwidth,clip=]{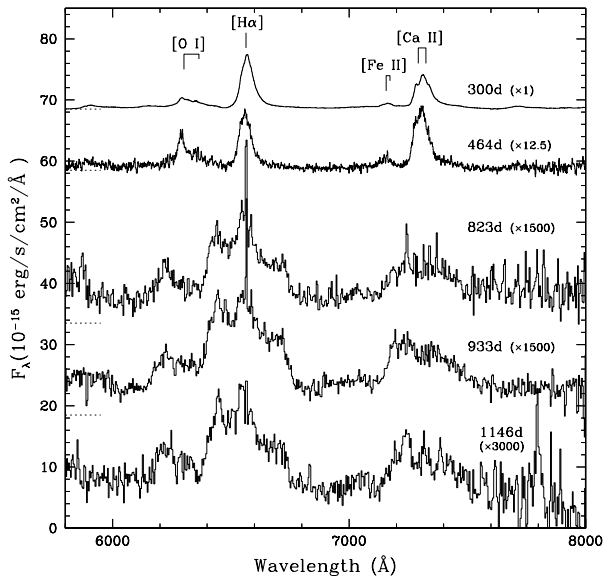}
\caption{Late-time optical spectra of SN~2004et.  The earlier two are
from \citet{sahu:06} (via the SUSPECT database).  The later three are
from the Keck telescopes. The approximate fluxes from two faint stars
lying within the seeing disk have been subtracted from the Keck
spectra (see text).  All the spectra have been scaled by the amounts
shown in brackets. In addition, the upper-four spectra have been
shifted vertically for clarity. The horizontal dotted lines indicate
the zero-flux levels for these spectra. \label{fig4}}
\end{centering}
\end{figure}

Due to the faintness of the supernova ($V \approx 22$ mag) 
the signal-to-noise ratio (S/N) is generally low. 
Spectra in the 5800--8000~\AA\ region are shown in Fig. \ref{fig4}, 
binned to a pixel width of 4~\AA\ ($\sim$180\,\kms), and compared 
with spectra obtained by \citet{sahu:06} on days~300 and 465. The 
characteristic ejecta H$\alpha$ emission profile seen in the \citet{sahu:06} 
spectra may be still apparent at 823 and 933~days, with a half width
at zero intensity (HWZI) of $\sim$2000\,\kms.
Its presence at 1146~days is less certain. However, on days~823, 933,
and 1146 a wide, steep-sided, box-like (``square'') component with
HWZI = 8500\,\kms\ is also present. Given the large decline in the
SN flux between day~465 and 823, it is difficult to determine
when the wide component actually appeared. A similar broadening of
the [Ca~II] feature at 7300~\AA\ can be seen. The [O~I] 
$\lambda\lambda$6300, 6364 line shows a shift to the blue of 
$\sim$3000\,\kms. We believe that this is due to the same 
``box-like profile'' phenomenon as seen in the
H$\alpha$ and [Ca~II] lines.  However, as the [O~I] line is
weaker and close to H$\alpha$, we only see the stronger blue
wing.  

Superimposed on the box-like profile of H$\alpha$ is a narrow component
that is clearly visible in all three Keck spectra. In the highest
resolution spectrum (day~933) the line is barely resolved at a
resolution of about 1.5~\AA\ (70\,\kms). However, the profile does
appear asymmetric, with a blue wing extending to about 3~\AA\
(140\,\kms) from the line peak.  Narrow [N~II] $\lambda\lambda$6548,
6583 emission is also clearly visible, especially in
the day~823 spectrum; a feature at 6717~\AA\ due to [S~II]
may also be present, although the corresponding [S~II] $\lambda$6731 
line and the [O~III] $\lambda\lambda$4959, 5007 lines are absent.

The presence of this narrow feature is intriguing.
While we cannot conclusively rule out a line-of-sight H~II region as
being responsible for the narrow emission, we note that the feature 
is not inconsistent with flash-ionized, undisturbed circumstellar 
material resulting from the progenitor wind. 
Also, between days~823 and 933, this component shifts by +90\,\kms;
currently, we have no explanation for this behavior, but note that 
it is not due to an error in the wavelength calibration which was
cross-checked with respect to the night-sky lines (see above).  
Furthermore, there is no report of any narrow features in a
high-resolution echelle spectrum taken about a week after discovery
\citep{zmm:04}

We defer further discussion of the circumstellar interaction to
\S\,\ref{sec:cf_idm}.

\section{Analysis}
\label{sec:analysis}
\subsection{Evidence for Dust}
\label{sec:mir_dust}
The evolution of the MIR spectral continuum and the spectral energy
distribution (SED) point strongly to IR emission from dust playing a
significant and increasing role in the overall flux distribution of
SN~2004et. To explore this proposition further, we make use of the MIR
photometry (see Table~\ref{tab:phot}) and spectra (Fig.~\ref{fig3}),
together with optical photometry \citep{sahu:06,misra:07} and spectra
\citep{sahu:06}\footnote{These were obtained from the SUSPECT
database: http://bruford.nhn.ou.edu/$~$suspect/index1.html.}.  In
addition, for the period 823--1146~days we derived approximate optical
magnitudes from the Keck spectra and from the study of
\citet{crockett:09}, as follows.

\citep{crockett:09} report {\it Hubble Space Telescope (HST)} WFPC2 
and NICMOS photometry of the SN~2004et field, using observations on 
day~1019 from program GO-11229 (PI: Meixner).  These show that the single 
point source at the SN location seen in ground-based images resolves into 
three point sources.  The brightest of these lies at the SN position and 
we assume that this source is, or at least contains, the actual
SN. They provide WFPC2 and NICMOS magnitudes for the
three sources. We have made use of these results to correct for the
flux of the two non-SN sources in the 823--1146~day optical data. To
convert to $BVRI$ magnitudes we first matched a double blackbody to
the combined fluxes of the non-SN sources. The resulting function was
multiplied by typical $BVRI$ transmission functions and then compared
with a similarly processed model spectrum of Vega.  The $BVRI$
magnitudes for the {\it HST} SN source (day~1019) were obtained similarly,
using a single-blackbody match. We estimate an uncertainty of $\pm$0.1
mag in the values derived by these procedures.  The {\it HST}-based
magnitudes are shown in Table \ref{tab:keckphot}.  

\citet{crockett:09} also report ground-based $BVRI$ photometry of 
SN~2004et on day~1054, obtained using the William Herschel Telescope 
(WHT) AUX imager.  We corrected these data for the
two nearby non-SN sources, and the net magnitudes are given in
Table \ref{tab:keckphot}.  

To derive $BVRI$ magnitudes from the Keck spectra, the non-SN
double-blackbody was first subtracted from each spectrum.  The net
spectra were then multiplied by $BVRI$ filter transmission functions
and compared with a similarly processed model spectrum of Vega.  The
values obtained are given in Table \ref{tab:keckphot}. (There is no
$B$ value on day~933 since at this epoch there was only partial
coverage of the relevant part of the spectrum.)  Given the difficult
observing conditions, we estimate that the magnitudes based on the
Keck spectra have a precision of $\sim\pm$0.2~mag.

\begin{figure}[!t]
\begin{centering}
\includegraphics[height=0.48\textwidth,clip=]{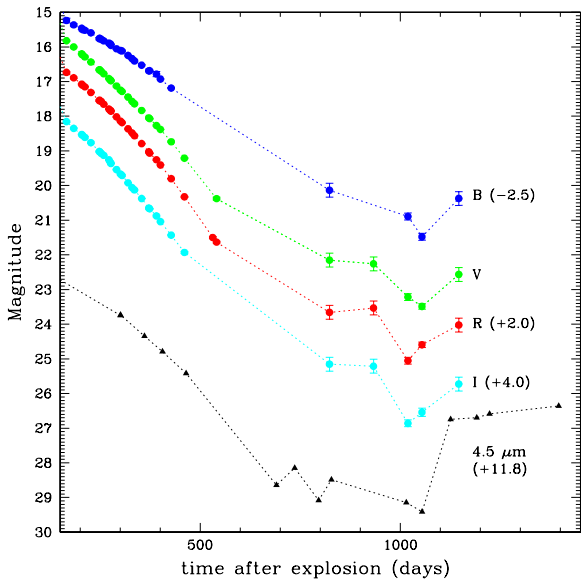}
\caption{Late-time optical light curves of SN~2004et. The pre-600~day
data are from Sahu et al. (2006) while the post-800~day magnitudes
were estimated from spectra taken at the Keck telescopes (see
\S~2.3), as well as from {\it HST} and WHT photometry \citep{crockett:09}
(see text and Table \ref{tab:keckphot}).  The light curves have been
corrected for the effects of two faint stars which lay within the
seeing disk.  Also shown for comparison is the 4.5~$\mu$m light curve.
For clarity, the $B$, $R$, $I$, and 4.5~$\mu$m plots have been shifted
vertically by the indicated amounts.
\label{fig5}}
\end{centering}
\end{figure}

In Fig. \ref{fig5} we show the later parts of the \citet{sahu:06}
light curves together with the 823--1146~day magnitudes described
above, all corrected for the fluxes from the two faint nearby stars.
The light curves appear to reach a minimum around $\sim$1000~days and
then begin to rise again. Similar behavior is seen in the MIR light
curves (Fig. \ref{fig2}); see also the 4.5~$\mu$m light curve
plotted in Fig. \ref{fig5} for comparison.  This will be discussed
later.

To take an initially neutral standpoint on the interpretation, we
compared blackbodies with the data, scaled to the IRAC epochs.  The
MIR spectra were scaled as explained above.  In addition, where
necessary, the light curves were used to scale the optical, PUI, and
MIPS data.  \\

\noindent {\it (a) Comparison with blackbodies at days~64--464:}\\ At
each IRAC epoch between 64~days and 464~days, blackbodies were matched
to the underlying optical and MIR spectral continua.  For the day~64
epoch it was found that a fair representation of the SED/spectral
continua required just a single, hot blackbody.  This is presumably
due to the ejecta photosphere dominating the flux at this phase. There
is little sign of thermal emission from dust.  We first considered the
match using $E(B-V) = 0.41 \pm 0.07$ mag \citep[][and references
therein]{misra:07}.  For a \citet{cardelli:89} law with $R_V=3.1$ this
corresponds to $A_V=1.27\pm0.22$ mag. However, we found that assuming 
the same extinction law, it was impossible to obtain a match to the
optical continuum without also significantly overproducing the MIR
flux. We found that $A_V=1.0$ mag provided a more satisfactory match and
so this value was adopted throughout the paper, using the
\citet{cardelli:89} law with $R_V=3.1$.

\begin{figure}[!t]
\begin{centering}
\includegraphics[height=0.48\textwidth,clip=]{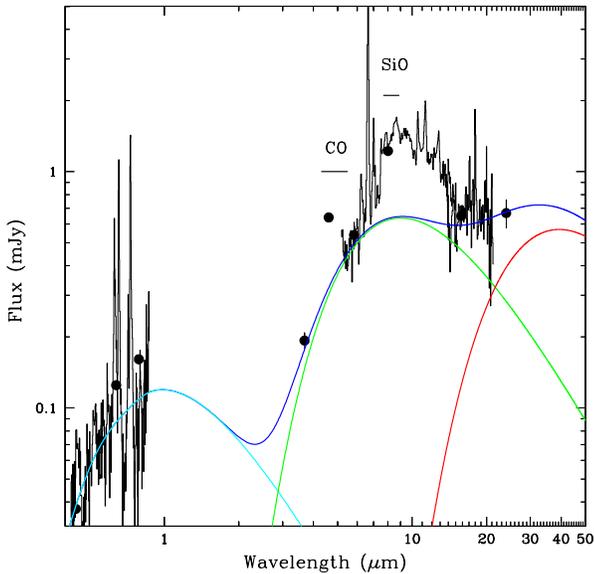}
\caption{Three-component blackbody match to the day~464 spectral
continua of SN~2004et.  The combined blackbody spectrum is shown in
blue while the hot, warm, and cold components alone are shown in cyan,
green, and red (respectively).  The blackbodies have been reddened
according to the \citet{cardelli:89} extinction law with $A_V = 1.0$
mag (see text). The optical spectrum is from \citet{sahu:06} and the
optical photometry from \citet{sahu:06} and \citet{misra:07}. The
optical, PUI, and MIPS fluxes have been interpolated to day~464.
\label{fig6}}
\end{centering}
\end{figure}

For the 300--464~day era it was found that a reasonable representation
of the spectral continua required three blackbodies:
(a) a hot (5000--10000\,K) blackbody to represent continuum 
optical emission from the hot ejecta, and (b) a combination of warm 
(450--700\,K) and cold (100--130\,K) blackbodies to 
represent the emission longward of $\sim$2.5~$\mu$m.  As an example, 
in Fig. \ref{fig6} we illustrate the contributions of these three
components for day~464. The blackbodies were matched to the
observations as follows.  The hot blackbody was first matched to the
continuum of the optical spectrum. The principal purpose here was to
estimate the effect that the hot continuum might have on the shorter
wavelength MIR fluxes.  In the $BV$ region the continuum dominated and
so the photometric points lay close to the continuum and the
model. However, in the $RI$ region strong emission from lines of
H$\alpha$, [O~I] $\lambda\lambda$6300, 6464, and [Ca~II] 
$\lambda\lambda$7291, 7323 meant that the photometric points lay 
above the continuum.

The warm and cold blackbodies were then matched to the MIR continuum.
During days~300--464, no attempt was made to reproduce the fluxes
around 4.5~$\mu$m or 8.0~$\mu$m, since these regions were strongly
affected by CO and SiO emission. Moreover, the broad emission in the
8--14~$\mu$m range could not be matched by a simple blackbody. (We
shall show below that this feature was due to silicate dust.)
Consequently, the warm/cold blackbody match was primarily to the
spectral continua around 6~$\mu$m and $>$14~$\mu$m plus photometry
points at 3.6~$\mu$m (IRAC), 16 and 22~$\mu$m (IRS peakup imager), and
24~$\mu$m (MIPS).  Iteration of the parameters of the three
blackbodies was carried out to optimize the overall match. \\

\noindent {\it (b) Comparison with blackbodies at days~690--795:}\\ 
For IRAC days~690--795, the only reasonably contemporary optical
spectrum was taken at the Keck telescope at day~823
(Fig. \ref{fig4}).  The optical spectrum comprises broad emission
lines on top of quasi-continuum emission. At IRAC day~795, for the
same reason as given above, the hot blackbody was adjusted to match
the continuum only. Thus, while a good match to the $V$ magnitude was
obtained, it underproduced the total fluxes in the $B$, $R$, and $I$
bands.  Given the indirect evidence (see below) that the broad
box-like profiles could have been influencing the spectrum as early as
$\sim$690~days, for epochs 690 and 736~days the hot blackbody
parameters were guided by the match at 795~days (i.e., the model was
adjusted to match the $V$-band flux but to underproduce in the $B$,
$R$, and $I$ bands).  Only one MIR spectrum was available (day~823, 
scaled to IRAC day~795), spanning 5--15~$\mu$m.  This showed little
evidence of strong CO/SiO/silicate emission, although the spectrum is
very noisy in the 5--7.5~$\mu$m region. We adjusted the warm/cold
models to match the spectral continuum, or the photometry fluxes where
there was no spectral coverage.

\begin{figure}[!t]
\begin{centering}
\includegraphics[height=0.48\textwidth,clip=]{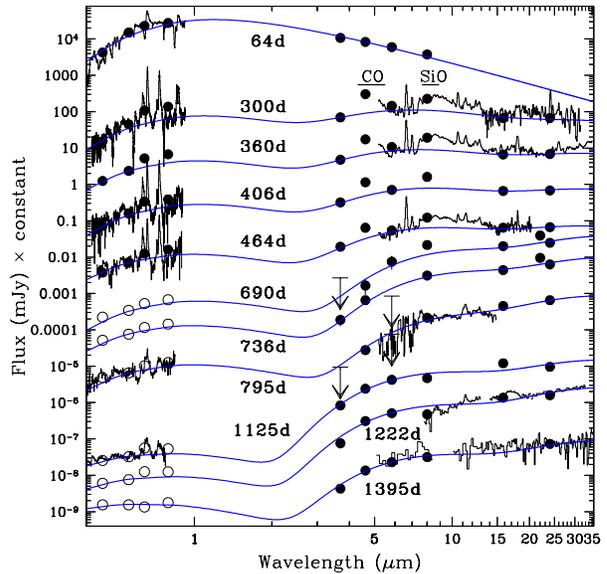}
\caption{One-component (day~64) and three-component (days~300--1395)
blackbodies (smooth continuous lines: see Table~\ref{tab:bb}) compared
with optical and MIR spectra/SEDs of SN 2004et.  The models have been
reddened according to the \citet{cardelli:89} extinction law with $A_V
= 1.0$ mag.  The photometry is indicated by filled circles.  The PUI,
MIPS, and spectral fluxes have been interpolated to the epochs of the
IRAC observations.  The open circles indicate estimated optical fluxes
obtained by interpolation or extrapolation of the light curves.  For
day~690 and later, the optical data have been corrected for the
effects of two faint stars which lay within the seeing disk (see
text).  The epochs are relative to the estimated explosion date of
2004 Sep. 22.0 \citep[][MJD = 53270.0]{li:05}.  The plots have been
shifted vertically for clarity.
\label{fig7}
}
\end{centering}
\end{figure}

We note that the 8$\mu$m flux on day~690 is underproduced by the model
(Fig.~\ref{fig7}), suggesting that the silicate emission, while
weaker, was still quite prominent at that epoch.  At subsequent epochs
the excess at 8~$\mu$m continued to fade. Indeed, by day~736, the
blackbody model provided quite a reasonable match in this wavelength
region, consistent with the near disappearance of the silicate feature
by then.\\

\noindent {\it (c) Comparison with blackbodies at days~1125, 1222 and
1395:}\\ For the final three epochs a similar procedure was
followed. At day~1125, the hot blackbody was matched to the 1146~day
optical spectrum scaled to day~1125. For day~1222 the warm/cold
blackbodies were matched to the day~1212 MIR spectrum scaled to
day~1222. These matches were then used to guide the choice of blackbody
parameters for, respectively, the Day~1125 MIR region and the day~1222
optical region. For the latest epoch the hot blackbody was matched to
optical fluxes obtained by extrapolation of the $BVRI$ light curves to
day~1395, while the warm/cold blackbodies were matched to the IRAC, MIPS
and IRS data from the period around day~1395.\\

The complete set of blackbody matches for days~64--1395 is shown in
Fig.~\ref{fig7}, and all of the blackbody parameters are listed in
Table~\ref{tab:bb}.  Also shown is the evolution of the blackbody
luminosities compared with the estimated total deposited radioactive
luminosity due to $^{56}$Ni, $^{57}$Ni, and other radioactive species,
assuming the deposition behavior specified by
\citet{li:93}.\footnote{$^{56}$Co decay dominates the luminosity
during most of the period studied, but by day~800 10\% is due to
$^{57}$Co decay and the contribution from this isotope grows with
time.}  (Note: Since the blackbodies were matched to the continua
only, their luminosities slightly underestimate the total
luminosities.)  Using the optical photometry around day~300
\citep{sahu:06} and the exponential tail method of \citet{hamuy:03}, we
estimate a $^{56}$Ni mass of $0.055\pm0.020$~M$_{\odot}$. This is
about 10\% less than the values found by \citet{sahu:06} and
\citet{misra:07} and is due to our reduced value of $A_V$, partially
compensated for by our slightly larger adopted distance.

On day~64 the hot component temperature was 5300~K and its luminosity
exceeded that of the radioactive decay deposition by a factor of 3.8.
These two points are consistent with emission from the
recombination-determined photosphere during the plateau phase.  Up to
day~795, the hot component remained around 5000--7000~K, but its
luminosity faded rapidly.  A rise in apparent temperature and
luminosity was seen on days~1222 and 1395. However, the optical fluxes
at these epochs were obtained by uncertain extrapolation of the
late-time light curves and so the blackbody parameters are
approximate, especially at the later eoch.

The warm component cooled and faded monotonically between days~300 and
795. In addition, the blackbody surface never exceeded
$\sim$1600\,\kms.  Furthermore, the warm blackbody luminosity was less
than or comparable to the radioactive input up to day~795.  This
behavior is consistent with the warm emission arising from newly
formed dust in the ejecta, and supports the optically based claims of
\citet{sahu:06} and \citet{misra:07} that ejecta dust formation took
place during this period.  By day~1125 the warm component luminosity
had {\it increased} by a factor of 3.3 since day~795 and exceeded the
radioactive deposition by a factor of over 60. In addition, its
temperature had risen slightly.  These points imply that an additional
energy source must have come into play.  This will be discussed
further in \S~3.2.  An interesting point is that the warm blackbody
radius remained constant to within $\pm17\%$ throughout days~300 to
1395. This will also be examined later.

The cold component maintained a roughly constant temperature of
$120\pm10$~K throughout the day~300--1385 era.  The velocity declined
monotonically from 12000\,\kms\ on day~300 to $\sim$6000\,\kms\ on
days~1125--1395.  The luminosity rose slowly during days~300--795, and
then more rapidly during the final three epochs.  The high velocities
immediately rule out an origin for the cold component in newly formed
ejecta dust.  In particular, these velocities are derived from
blackbody matches.  Consequently the velocities correspond to the
smallest possible source radii consistent with the cool component
SED. Yet the required velocities are still too large for the source to
be attributable to ejecta dust.  Also, again since these are blackbody
matches, these results are independent of the nature of the dust.
Furthermore, there is currently no known mechanism by which newly
formed ejecta dust rapidly cools, leaving behind only a small warm
component that is amenable to detection at MIR wavelengths.  Nor is it
clear that such a scenario could be made to be consistent with the
optical depths indicated by the MIR spectra. We propose instead that,
at least up to day~795, the cold component arose from dust whose
existence preceded the supernova explosion.  The obvious mechanism for
this radiation is an IR echo.

\subsection{More Detailed Interpretation of the IR Emission}
\label{sec:interp}

Guided by the results of the blackbody study above, we carried out a
more detailed investigation of the three components of the SN spectrum.
Regarding the hot component, \citet{wooden:93} showed that during the
second year of SN~1987A, the dust-emission continuum could be
contaminated by blackbody emission from hot, optically thick gas, as
well as by free-bound radiation. Here we represent both effects using
a single hot blackbody, as before.  The hot blackbody was adjusted to
match the underlying continuum of the optical spectrum as discussed
above.  The free parameters were blackbody radius (usually introduced
as epoch and velocity) and temperature.

For the warm component, we note from the blackbody study that in the
period 300--795~days, the IR emission is consistent with an origin in
newly condensed ejecta dust, heated by radioactive decay. However, an
alternative source could be dust in the progenitor wind (i.e., an IR
echo), and we first examine this possibility. Comparison with IR echo
models \citep[][]{meikle:06} suggests that the warm component
could indeed be due to reradiation of the SN flux from pre-existing
circumstellar dust.  For a dust-free cavity formed by evaporation by
the SN peak flux, over a range of CSM density profiles ($n = -1$ to
$-2.3$), we find that the echo model light curve actually declines too
rapidly.  However, if we increase the cavity size by a factor of $\sim 
10$, then the IR echo model can reproduce the characteristic dust
temperature and SED decline over the 300--795~days period. Such a large
cavity would have to be the result of the episodic mass-loss history
of the progenitor.  

The problem with this scenario is the {\it ad hoc}
nature of the ``fine tuning'' of the cavity size required to make it
work. There is no independent evidence to support the occurrence of a
mass-loss event which yielded a cavity of just the right size to
account for the warm-component behavior and support the
IR~echo scenario. On the contrary, \citet{sahu:06} argue that line
shifts to the blue in their late-time optical spectra imply dust
formation in the ejecta.  We have independently reexamined their
spectra.  In the [O~I]~$\lambda$6300 line we find that during
days~314--465 the red wing shifts by about 800\,\kms\ to the blue
while the blue wing is unmoved.  We therefore agree with
\citet{sahu:06} that this is evidence of dust condensation in the
ejecta.\footnote{In the H$\alpha$ profiles, also presented by
\citet{sahu:06}, between days~301 and 314 we find a jump of
$-400$\,\kms\ in the whole profile, but little sign of a progressive
blueshift in the subsequent day~314--465 period. During this period
the shift remained at about $-165$\,\kms\ (corrected for the redshift
of NGC~6946). We suggest that this could simply be due to the peculiar
velocity of the supernova center of mass.  We have no explanation for
the sudden day~301--314 jump.  Nor do we see the ``flattening'' in the
profile reported by \citet{sahu:06}.  We conclude that there is little
evidence of a progressive blueshift in the H$\alpha$ line.  However,
given the highly extended nature of the hydrogen envelope, dust formed
in the much smaller refractory-element zone of the SN would be
unlikely to produce a significant modification of the profile, except
perhaps a truncation of the extreme red wing.}

In addition, \citet{sahu:06} and \citet{misra:07} found that in the
period 310--370~days the rate of decline in the optical light curves
accelerated. Such behavior is also suggestive of dust condensation in
the ejecta, although alternative explanations for the steepening are
(a) increasing transparency of the expanding ejecta to the gamma rays
from the radioactive decay, or (b) the onset of the IR
catastrophe. Nevertheless, taken with the [O~I]~$\lambda$6300 blueshift
and the necessarily specific selection of the cavity radius for the 
IR-echo model to work, we conclude that the warm component during
days~300--795 was most likely due to newly formed,
radioactively-heated dust in the ejecta.  We also note that no
detectable IR echo would be expected from a dusty cool dense shell
(CDS; see below), since the CDS would not have formed until long after
the bulk of the SN UV/optical flux had passed. \\

\subsubsection{The Warm Isothermal Ejecta Dust Model} 
\label{sec:idm}
We investigated the ejecta-dust hypothesis using a simple analytical
IR-emission model \citep{meikle:07} comprising a uniform sphere of
isothermal grains with the luminosity and SED obtained via the escape
probability formalism \citep{ost:89,lucy:89}.  At wavelengths beyond
$\sim$14~$\mu$m we found that, as with the blackbody study, the warm
isothermal dust model (IDM) increasingly underproduced the observed
flux. Consequently, a cold component had to be included.  Similar to
the earlier blackbody study, we found that for the period
300--795~days the cold-component flux could be reproduced with a
blackbody at $\sim$110~K, radius $(3.0-5.5)\times10^{16}$~cm, and
luminosity $(2-3) \times 10^{38}$~ergs s$^{-1}$. The source of this radiation
cannot be newly formed dust in the SN ejecta. To attain the required
blackbody radius at a given epoch would require velocities as high as
12000\,\kms, far greater than the velocities of the fastest
moving refractory elements.  We therefore invoked an IR echo from
pre-existing dust to account for the cold component. This was based on
the IR~echo model of \citet{meikle:06}, and will be described in
detail below.

To select the likely grain density distribution and grain materials
for the IDM, we were guided by dust condensation calculations and the
explosion models upon which they are based. Only a few papers have
been published which describe local SN dust condensation based on
explosion models.  \citet{kozasa:89} and \citet{todini:01} have
calculated dust condensation within the ejecta of SN~1987A, while
\citet{bianchi:07} have examined dust formation in supernovae having
progenitors of 12--40~M$_{\odot}$ and solar metallicity.  These
authors used the ejecta chemical composition as determined in
nucleosynthesis models \citep{nomoto:91, woosley:95}. All adopted
complete chemical mixing within the dust-forming zone. Within this
zone \citet{todini:01} and \citet{bianchi:07} assumed a uniform
density distribution, while \citet{kozasa:89} used the density profile
from an explosion model \citep{hashimoto:89}, but this is also roughly
flat. Dust-type abundances were determined by all the authors, but
none made explicit predictions about the dust distribution within the
ejecta.  Three-dimensional core-collapse SN explosion models 
\citep{kifonidis:06} confirm that extensive mixing of the 
core takes place. In addition, the same models
show that the density structure is likely to be exceedingly
complex, with high-density clumps moving out through lower-density
gas. Qualitatively similar results are reported for jet-driven
explosions of red supergiants by \citet{couch:09}.
How this affects the dust distribution has yet to be determined.

Given the current state of knowledge, we assume that dust of uniform
number density forms throughout the zone containing abundant
refractory elements. The extent of this zone can be assessed using the
late-time widths of metal lines. In the nebular optical spectra of
SN~2004et \citep{sahu:06} the maximum velocities implied by the metal
lines generally do not exceed $\sim$2500\,\kms.  This upper limit is
adopted as the size of the dust-forming region. The uniform-density
assumption is conservative in that it provides the least effective way
of hiding dust grains in optically thick regions. We initially
considered both silicate and amorphous carbon dust, with the mass
absorption functions taken from \citet{laor:93} and
\citet{rouleau:91}, respectively.  

The IDM comprises a uniform sphere of isothermal dust grains.
Following the escape probability formalism \citep{ost:89,lucy:89}, the
luminosity ($L(\nu)$) of the sphere at frequency $\nu$ is given by
$L(\nu)=4\pi^2R^2B(\nu,T)[0.5\tau(\nu)^{-2}(2\tau(\nu)^2-1+(2\tau(\nu)+1)e^{-2\tau(\nu)})]$,
where $R$ is the radius of the dust sphere at some time after the
explosion, $B(\nu,T)$ is the Planck function at temperature $T$, and
$\tau(\nu)$ is the optical depth to the center at frequency $\nu$.
For a grain-size distribution $dn =ka^{-m}da$, where $dn$ is the
number density of grains having radius $a \to a+da$, $m$ is typically
between 2 and 4, and $k$ is the grain number density scaling factor,
it can be shown that $\tau(\nu)=\frac{4}{3}\pi
k\rho\kappa(\nu)R\frac{1}{4-m}[a^{4-m}_{(max)}-a^{4-m}_{(min)}]$,
where $\rho$ and $\kappa(\nu)$ are, respectively, the density and mass
absorption coefficient of the grain material.  The grain-size
distribution law was set at $m=3.5$ \citep{mathis:77} with
$a_{(min)}=0.005~\mu$m and $a_{(max)}=0.05~\mu$m.  The total mass of
dust, $M$, was then found from $M=4\pi R^2\tau(\nu)/3\kappa(\nu)$
\citep{lucy:89}. For a given grain material (silicate or amorphous
carbon) the free parameters were grain temperature and grain number
density scaling factor, $k$.

\subsubsection{Comparison of the IDM with the Observations} 
\label{sec:cf_idm}

We found that to achieve reasonable matches to the data it was
necessary to increase the dust mass until it was optically thick in
the MIR.  To make use of the much greater information available in the
MIR spectra (compared with the photometry), we first carried out model
matches to those epochs for which we had both MIR spectra and
reasonably contemporary photometry (see above). This comprises six
epochs, corresponding to IRAC days~300, 360, 464, 795, 1222 and
1395. The parameters of the hot blackbody, warm IDM, and cold IR echo
components were adjusted to optimize the overall match to the
continua.  Guided by these results, model matches were carried out for
the remaining epochs where no MIR spectra were available. We examined
the warm-model matches using silicate and amorphous carbon grain
materials.

\begin{figure*}[!t]
\begin{centering}
\includegraphics[height=0.85\textwidth,clip=]{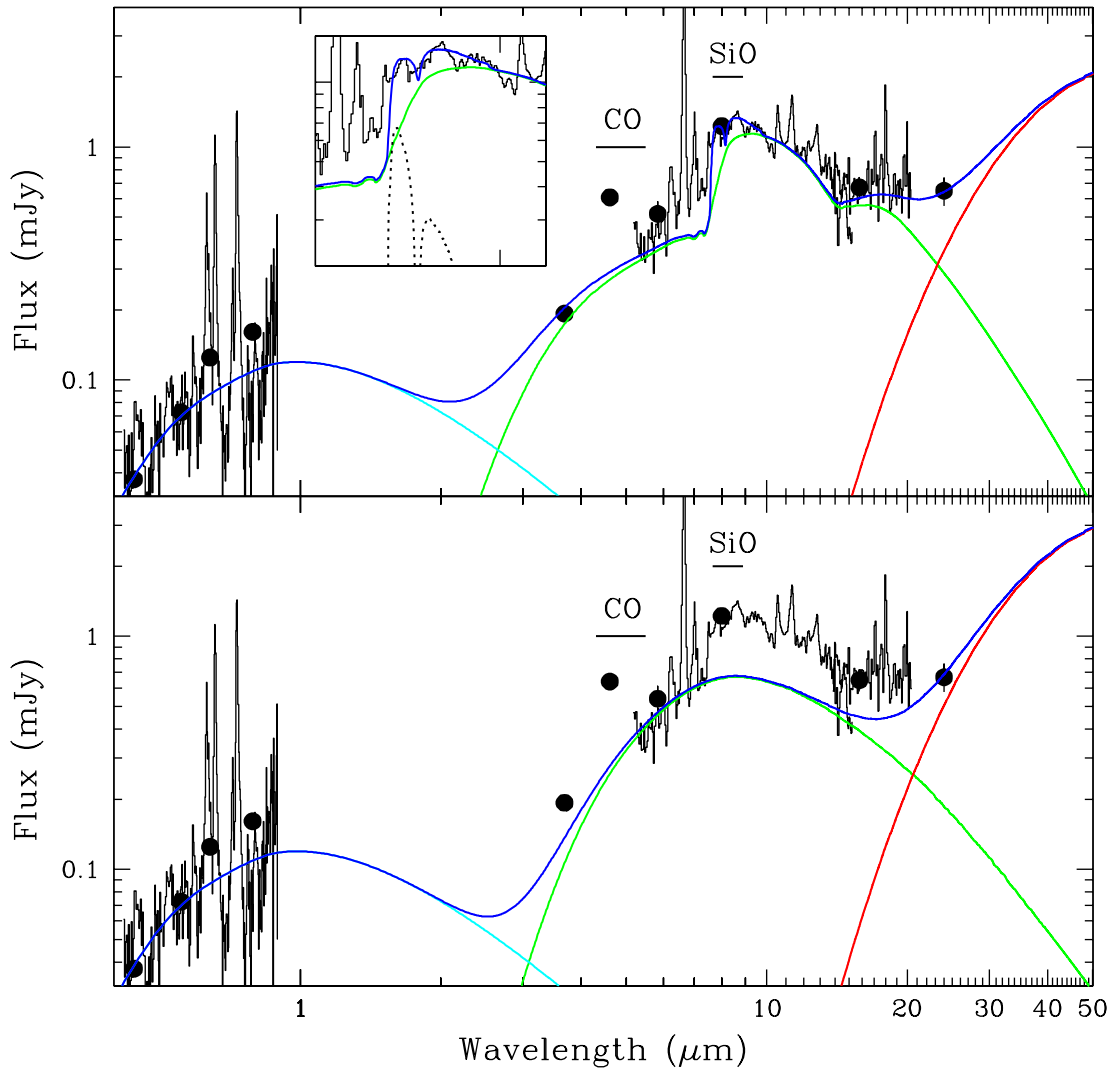}
\caption{Day~464 observations (black) of SN~2004et compared with
models. The upper (main) and lower panels show the IDMs (green) for
silicate and amorphous carbon grains, respectively. The total model
spectrum (blue) also comprises hot (blackbody: cyan) and cold
(interstellar IR echo: red) components.  The upper-panel model also
contains a contribution from the SiO fundamental. The inset shows the
separate SiO contribution (dotted line).  It can be seen that a
superior match to the spectrum is achieved with the combined silicate
dust and SiO model, as compared with the amorphous carbon dust model.
\label{fig8}
}
\end{centering}
\end{figure*}

\noindent {\it (a) 300--464~days:}\\ 
Inspection of the day~300, 360, and 464 MIR spectra suggests the
presence of SiO fundamental emission in the 7.7--9.5~$\mu$m region.
SiO was discovered in the Type~II-pec SN~1987A \citep{roche:91,
wooden:93} and more recently in the Type~II-P SN 2005af
\citep{kotak:06}.  To assess the contribution of SiO in SN~2004et we
added an SiO component to the silicate model spectrum.  The SiO models
were taken from a study of SiO in SN~1987A by \citet{ld:94}.  We used
their non-LTE (local thermodynamic equilibrium) 
models whose epochs lay most closely in time to the
above three SN~2004et observation epochs in the day~300--464 period, and
SiO masses were estimated assuming similar excitation conditions in
the two SNe.  The coexistence of SiO and silicates is not surprising
since SiO formation is an essential step in the silicate-formation
sequence \citep[e.g.][]{todini:01}. No SiO was added to the amorphous
carbon version of our model.

\begin{figure}[!t]
\begin{centering}
\includegraphics[height=0.48\textwidth,clip=]{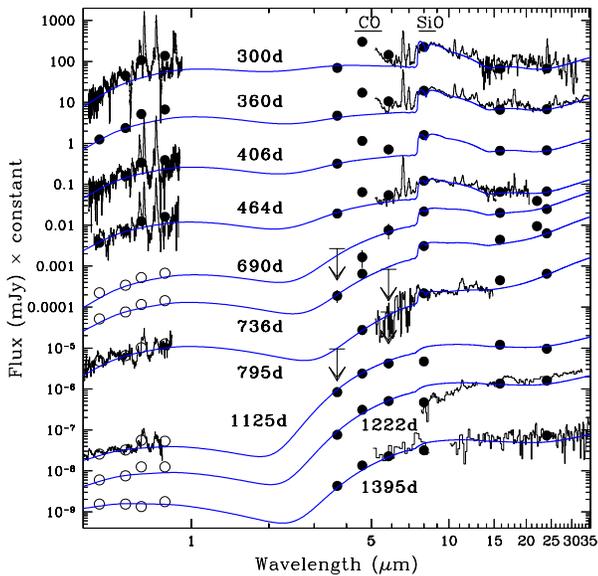}
\caption{Silicate dust models (smooth continuous lines) compared with
MIR SEDs of SN 2004et (see Table~\ref{tab:sio}). The models also
comprise hot (blackbody) and cold (interstellar IR echo) components
(see text).  The models have been reddened according to the
\citet{cardelli:89} extinction law with $A_V = 1.0$ mag. The
photometry is indicated by filled circles.  The PUI, MIPS, and
spectral fluxes have been interpolated to the epochs of the IRAC
observations. The open circles indicate estimated optical fluxes
obtained by interpolation or extrapolation of the light curves.  For
day~690 and later, the optical data have been corrected for the
effects of two faint stars which lay within the seeing disk (see
text).  The epochs are relative to the estimated explosion date of
2004 Sep. 22.0 (MJD = 53270.0) \citep{li:05}.  The plots have been
shifted vertically for clarity.
\label{fig9}}
\end{centering}
\end{figure}

In order to reproduce the fading visibility of the silicate feature
during the 300--464~day period, it was necessary to increase the
optical depth. We found that $\sim0.5\times10^{-4}\,{\rm M}_{\odot}$ of
silicate grains plus a few $\times 10^{-4}\,{\rm M}_{\odot}$ of SiO provided
good matches to the MIR continua.  The carbon models also required
$\sim10^{-4}\,{\rm M}_{\odot}$ of dust, but they produced much inferior
matches to the continua, especially in the 8--14~$\mu$m region.
Examples of matches at day~464 for the two grain materials are shown
in Fig. \ref{fig8}. All of the three-component model matches for silicate
grains are shown in Fig. \ref{fig9}, and the silicate model parameters
are summarized in Table~\ref{tab:sio}.

The success of the silicate model in reproducing the 8--14~$\mu$m
feature during days~300--464 provides strong support for the
newly formed silicate dust scenario, at least for this period.  The
dust temperature fell from $900^{+100}_{-140}$~K to
$650^{+50}_{-50}$~K. Even as early as day~300, the optical depth to
the center at 10~$\mu$m ($\tau_{10\mu m}$) had to be set as high as
2.8 in order to reduce the silicate feature to the observed
visibility. By day~464 the optical depth had increased to 3.6. Also,
the dust mass had grown to
$(0.66^{+0.27}_{-0.15})\times10^{-4}$~M$_{\odot}$.  

An interesting point about the mass estimates is that the silicate 
feature yields an additional constraint on the model at each epoch 
since the optical depth had to be adjusted to match the visibility 
of the observed 8--14~$\mu$m feature. Consequently, in spite of the high 
optical depth, within the uniformity limitation of the model the derived 
dust masses are actual values rather than just lower limits. Moreover,
since the optical depth, $\tau(\nu)$, is fixed by the silicate feature
visibility and the dust mass $M \propto \tau(\nu)$ (see above), it
follows that the dust mass is insensitive to the grain-size limits.
We also stress that in the silicate dust model, most of the observed
broad 7.5--14~$\mu$m emission feature is produced by silicate. SiO
makes a relatively minor contribution (see Fig. \ref{fig8}).
Nevertheless, the SiO mass was more than adequate to supply the
material for the newly formed silicate dust. This supports the
newly formed silicate grain hypothesis. The SiO decline from about
6$\times10^{-4}\,{\rm M}_{\odot}$ to 3.5$\times10^{-4}\,{\rm M}_{\odot}$ 
over the period may also suggest that it was indeed supplying the silicate
material.  However, it is possible that some of the mass decline was
not real but was actually due to reduced excitation as the ejecta
cooled from 900~K to 650~K during this period.  

\noindent {\it (b) 690--795~days:}\\ While we conclude that, up to at
least day~464, the newly formed grains in the ejecta of SN~2004et
were dominated by silicate material, the situation at later epochs is
somewhat less clear.  From day~690 onwards the visibility of the
8--14~$\mu$m silicate feature faded quite rapidly.  Indeed, by
day~795 there may only be a small excess remaining in the silicate
region.  The fading of the silicate feature could be due to (a)
increasing optical depth as more silicate dust forms, or (b) an
increasing contribution of IR emission from non-silicate dust in a
cool dense shell (CDS) formed by an ejecta/CSM collision. A third
possibility is the suppression of the silicate feature by the
increased prominence of emission from undisturbed non-silicate CSM
dust.  However, we have already dismissed above the likelihood of a
{\it warm} IR echo, so this mechanism is not considered further.
From as early as day~690, the hot+warm component increasingly
exceeded the radioactive input luminosity. Indeed, since the 
models were matched to the continua only, the true hot+warm
luminosities are even a little larger than those shown in
Table~\ref{tab:sio}.  The growing luminosity excess relative to the
radioactive input implies that an additional energy source must have
come into play by day~690. A possible explanation for the MIR
brightening is IR emission from dust produced in a shock-formed CDS due
to ejecta/CSM collision (see below).  Such a phenomenon was invoked to
account for IR emission from SN~1998S \citep{pozzo:04} and SN~2006jc
\citep{smith:08, mattila:08}.

An ejecta/CSM collision would also account for the appearance of the
wide, box-like (``square'') emission-line profiles in the optical 
region as well as the rebrightening of the optical light curves. As
mentioned above, given the large decline in the supernova optical flux
between days~465 and 823 it is difficult to determine when the wide
components actually appeared.  However, given that we see the light
curve leveling off by $\sim$800~days, it seems likely that the box-like
profiles emerged at about the same time.  Similar box-like profiles
were seen in the H and He lines of the Type~IIn SN~1998S
albeit with a much earlier onset
\citep{gerardy:00,leonard:00,pozzo:04}, as well as in the Type IIb
SN~1993J \citep{matheson:00,fransson:05}.  For SN~1998S, \citet{gerardy:00}
proposed that the box-like profiles were due to an ejecta/CSM collision.
We suggest that here, too, the appearance of the box-like profiles in
SN~2004et was produced by the impact of the fastest moving ejecta on a
pre-existing CSM.  From the epoch and velocity of the wide H$\alpha$
component, we can infer a shell of radius $8\times10^{16}$~cm or
5600~AU at day~1146. For a red supergiant wind of velocity 10\,\kms\,
this would indicate a CSM age of $\sim$2700~yr. 

The appearance of the boxy profiles of H$\alpha$, [Ca~II], and
[O~I] by day~800 puts constraints on the density of the
circumstellar environment of SN~2004et. A requirement for a
CDS to form behind the reverse shock of the supernova is that
the cooling time of the shock is shorter than the adiabatic time
scale. The cooling time can be estimated as $t_c = 4.6 \times 10^{-3}
\dot M_{-5}/u_{\rm w1}^{-1} V_{\rm s4}^{5.34} t_{\rm days}^2$ days
\citep{chevfrans:03}, which is valid for a power-law ejecta
with $\rho \propto V^{-11.7}$, typical of a red supergiant progenitor.
The requirement that the reverse shock should be radiative until at
least day~1150, as indicated by the last spectrum, and using $V_s
\approx 10^4 \rm ~ km ~s^{-1}$, translates into a requirement that
$\dot M \ga 2 \times 10^{-6} \rm ~ M_\odot\ yr^{-1}$ for $u_w = 10 \rm
~ km~ s^{-1}$. This is in agreement with both the value of $\dot M
\approx 1\times 10^{-5} (T_{\rm CSM}/10^5 {\rm K})^{0.75} \rm ~ M_\odot
yr^{-1}$ from the radio observations \citep{chev:06}, and from
the X-ray observations, $(2-2.5)\times10^{-6} \rm ~ M_\odot yr^{-1}$
\citep{rho:07}. The former depends on the temperature of the
circumstellar medium, $T_{\rm CSM}$, which is likely to be in the range
$(2-10) \times 10^4$ K, resulting in at least the same mass-loss 
rate as the X-ray observations. We therefore conclude that the
appearance of the boxy line profiles from the CDS is fully consistent
with previous mass-loss determinations from this object. It is,
however, interesting that SN~2004et is the first SN~II-P for
which this has been observed. 

The question now arises: when did the CDS dust begin to significantly 
affect the MIR emission?  Between days~300 and 795 the IDM
luminosity (Table~\ref{tab:sio}) declined roughly exponentially with a
$\sim$200~day e-folding time. There is little sign of an increase in
this time constant even as late as day~795, such as one might expect
if CDS dust emission was becoming important.  We also note that even
as late as day~736 the IDM luminosity was less than that of the
radioactive deposition, although by day~795 it exceeded it by $\sim$25\%.  
We therefore suggest that, while the ejecta/CSM impact probably
began as early as day~690, significant emission from CDS dust began
somewhat later, at about day~800. 

Thus, we decided that the IDM could provide useful estimates of the
ejecta dust mass up to day~795.  We matched the warm component
allowing the silicate feature visibility to take as large a value as
was consistent with the data.  Up to day~795 we were still able to
obtain fair matches to the data.  The apparent decline in the mass of
SiO continued, suggesting the ongoing removal of the molecules to form
silicate grains. However, as in the day~300--464 period, some of the
decline in the SiO feature may have been due to ejecta
cooling. Compared with earlier epochs, the temperature declined more
slowly, reaching $400^{+20}_{-40}$~K, and the 10~$\mu$m optical depth
increased more rapidly, reaching 11.5. The dust mass also increased
more rapidly, reaching $(1.5^{+1.2}_{-0.4})\times10^{-4}$~M$_{\odot}$
at day~795.  Given the weakness of the silicate feature by this epoch,
our value should really be treated as an upper limit. However, it
should also be kept in mind that, by this time, some of the MIR flux
may have been due to the CDS dust, reducing the amount of directly
observed ejecta dust that is inferred.

\noindent {\it (c) 1125, 1222 and 1395~days:}\\ Between days~795 and
1125 the IDM luminosity increased by a factor of 5, exceeding the
radioactive deposition by a factor of $\sim$100.  Nevertheless, for
completeness, we continued the IDM matching in the day~1125--1395 era.
On days~1125 and 1222, the model used above provided a much less
satisfactory fit to the SED (Fig. \ref{fig9}). In particular, the
model now overproduced in the 8~$\mu$m region.  Indeed, there is some
evidence that the previously observed silicate emission feature
evolved into an absorption component. The model could not reproduce
this effect. However, by day~1395 the ``absorption'' had largely
disappeared and a better model match was obtained. Nevertheless we
conclude that by day~1125 the uniform-density ejecta dust-emission
model was no longer appropriate.  In any case, as already indicated,
by this epoch the vast bulk of the warm-component flux could not have
been due to new, radioactively heated ejecta dust, and was probably
due instead to CDS dust.

The optical spectra also provide evidence for the presence of
CDS dust.  The tops of the box-like profiles in the day~823--1146
optical spectra (Fig. \ref{fig4}) exhibit a decline from the blue to
the red suggesting the presence of dust attenuation within the
emission region.  The large, high velocity widths of the boxy
features in SN~2004et immediately rule out new ejecta dust as being
the main source of the attenuation.  Fading across the H and He
profiles was also seen in SN~1998S. \citet{pozzo:04} showed that the
attenuation was probably due to dust which condensed in a CDS formed
behind the reverse shock.  We conclude, therefore, that the later MIR
rebrightening was due to IR emission from dust formed in the CDS.  As
the ejecta shock moved into the CSM, a CDS was formed where dust could
condense, resulting in the blue-to-red fading in the line profile and
producing the excess infrared flux.

To make a rough estimate of the CDS dust mass present during the
day~1125--1395 period, we applied the IDM to a shell of dust lying at
6000~AU. Adjusting the model to reproduce the MIR SEDs suggests CDS
dust masses of $\sim 5\times10^{-4}\,{\rm M}_{\odot}$ (amorphous
carbon) or $\sim 2\times10^{-4}\,{\rm M}_{\odot}$ (silicates).  Such
masses of the required refractory elements could be produced by a
progenitor wind.  The corresponding temperatures and optical depths at
10~$\mu$m are 330~K and $\sim$0.03 for amorphous carbon and 500~K and
$\sim$0.025 for silicates. The corresponding $V$-band optical depths
are 0.6 (amorphous carb.) and 0.015 (silicates). These are in addition
to the $A_V=1.0$ mag which was present before the CDS formation.  We
note that while the CDS model can provide a fair representation of the
fluxes over most of the MIR range covered, in the 8--14~$\mu$m region
the observed ``dip'' is not reproduced. Indeed, the silicate match is
very poor in this region owing to the strong silicate emission
feature.  The observed dip may be a sign that the dust in the CDS has
actually condensed into optically thick clumps which, if silicate in
composition, might produce a broad absorption in this region. The CDS
dust masses for SN~2004et are about a factor of 10 less than the lower
limit for the CDS dust mass obtained by \citet{pozzo:04} for the
Type~IIn SN~1998S.  However, at about $8\times10^{16}$~cm, the radius
of the SN~2004et CDS is much larger than the $\sim1.5\times10^{16}$~cm
CDS radius in the case of SN~1998S.  Further investigation of the CDS
is beyond the scope of this paper.

\subsubsection{Dust Clumping}
An intriguing result from both the warm blackbody models and IDMs is
the remarkable constancy (standard deviation $<10\%$) of the model 
radii throughout the day~300--795 era 
(Tables~\ref{tab:bb},\ref{tab:sio}), during which time
the SN itself expanded by a factor of $795/300 = 2.65$.  How
might this occur? Given the assumed uniform density, the high optical
depth throughout the period implies that the MIR ``photosphere'' would
always lie quite close to the physical surface of the radiating dust
extent of the model.  It is therefore conceivable that, if the
photospheric radius remained fixed, as the ejecta flowed through it
the model radius would also remain roughly constant. But for this
scenario to work the dust flowing beyond the photosphere would have to
be replaced within the photosphere at just the right rate to maintain
the fixed photospheric radius.  This seems to be a rather contrived
scenario.  A more plausible explanation is that the dust actually
formed in small, dense, comoving clumps. Such clumps might arise in
localized regions of density enhancement.  With more efficient cooling
in such regions, the clumps might be unable to expand against the
hotter, higher pressure interclump gas.  We know that in the case of
SN~1987A there is strong evidence that the dust formed in clumps
\citep{lucy:91}.

In order to explore the effects of clumping on the MIR SED, we
extended the silicate IDM so that it comprised clumps of dust
grains, and calculated the SEDs for days~300--795.  We
considered a spherical ``cloud'' of spherical clumps, with a cloud
radius of $R_{\rm cd}$. The clumps are uniformly distributed within the
cloud. At any given time, each clump has the same uniform number
density of dust grains.  The free parameters of the grains are the
same as for the IDM: $m$, $k$, $a_{\rm (max)}$, and $a_{\rm (min)}$.  
For this study, the grain number density is allowed to vary, but $m$,
$a_{\rm (max)}$, and $a_{\rm (min)}$ are fixed at the same values as 
adopted for the IDM described earlier.  The SED was then derived using 
a treatment analogous to that of the uniform IDM.  A clump-size
distribution, $dn_{\rm cl}=k_1R_{\rm cl}^{-m_1}dR_{\rm cl}$ is invoked, 
where $dn_{\rm cl}$ is the number density of clumps having radius 
$R_{\rm cl} \to R_{\rm cl}+dR_{\rm cl}$, and $k_1$ is the clump 
number density scaling factor (i.e., it controls the total number 
of clumps in the cloud, $N_{\rm cl}$). Thus, the free parameters 
of the clumps are $m_1$, $R_{\rm cl(max)}$, $R_{\rm cl(min)}$, 
and $N_{\rm cl}$.  The model is restricted to the case of
nonmerging clumps.  The cloud radius is fixed at $v \times t$ where
$v=2500~\kms$, this being the extent of refractory elements indicated
by the metal lines (see above). The SiO contributions were assumed to
be the same as in the IDM case.  For the hot and cold components we
used the same hot blackbody and echo model results as before.  Details
of the cloud of clumps model are given in the Appendix. 

The cloud of clumps model was matched to the data using the same
criteria as for the IDM case.  We considered a simple case where
$m_1=0$ (a flat number density distribution of clumps). For a given
$N_{\rm cl}$, the model match to the SED is obtained by adjusting
$R_{\rm cl(max)}$ and the grain number density.  This match can be
maintained over a wide range of $N_{\rm cl}$ by adjusting the clump radius
and dust density so that the total dust mass is unchanged. In this
study we fixed $N_{\rm cl} = 2500$ for all epochs.  The total dust mass was
also insensitive to the choice of $R_{\rm cl(min)}/R_{\rm cl(max)}$, and 
this was set at 0.01.  Using the same model temperatures as for the IDM,
virtually identical matches to those of the IDM were
obtained.  The cloud of clumps parameters are listed in
Table~\ref{tab:clo}.

As expected, there was little change in clump radii with time.
$R_{\rm cl(max)}$ first rose to 15\% above and then fell to $\sim20$\%
below the value at 300~days. The filling factor, $f$, defined as the
ratio of the volume of clumps to that of the cloud, fell monotonically
from $1.4\times10^{-2}$ to $0.044\times10^{-2}$.  The clump optical
depth evolution at 10~$\mu$m increased almost monotonically with time,
with the largest clumps closely matching the high optical depths of
the IDM. In contrast, the optical depth of the cloud decreased from
0.4 to 0.05. As with the single-sphere IDM, the need to reproduce the
silicate feature placed strong constraints on the model. For example,
if we decreased the clump optical depths, and at the same time
increased the cloud optical depth to maintain the overall level of the
SED, the silicate feature was increasingly suppressed below the
observed visibility. The reverse procedure produced a silicate feature
that was too strong.  The total dust mass was very similar to that of
the IDM, increasing monotonically from $0.4\times10^{-4}\,{\rm
M}_{\odot}$ to $1.4\times10^{-4}\,{\rm M}_{\odot}$. These results are
as would be expected for a scenario where dust continues to form in
clumps of fixed size and where the clumps are comoving with the
expansion of the supernova.

It has been believed for some time that clumping occurs in SN ejecta
\citep[e.g.,][]{lucy:91,hachisu:91,herant:94}.  Moreover
\citet{fesen:01} has directly observed small knots [$(1-2)\times 10^{-4}$
of the extent of the dust-rich ejecta] in the SNR Cassiopeia~A.  However,
the size distribution and evolution of such clumps is not well
understood and does not allow us to constrain the clump model for
SN~2004et.  Nevertheless, at least for a flat number distribution of
uniform-density clumps, we find that the introduction of clumping has
a negligible effect on the estimated mass of dust. In particular, for
the 300--465~days period, the derived masses are actual values rather
than upper limits.  It may be that by introducing nonuniform
densities within the clumps, larger dust masses could be consistent
with the observations, but such calculations are beyond the scope of 
this paper. Nevertheless, this brief study of the effects of 
clumping adds to the growing weight of evidence that the mass of 
grains produced in supernova ejecta can be only a minor contributor 
to the total mass of cosmic dust.

\subsubsection{Interpretation of the Cold Infrared Source}
\label{sec:cold}

To investigate the possibility that the cold component was due to an
IR echo from pre-explosion dust, we used an IR-echo model similar to
that described by \citet{meikle:06}. The model follows those of
\citet{be:80}, \citet{wright:80}, \citet{dwek:83}, and
\citet{graham:86}.  The input bolometric light curve is a parametrized
version of the $UBVRI$ ``bolometric'' light curve (BLC) of
\citet{sahu:06}: $L_{\rm bol} = L_0 {\rm exp}(-t/\tau)$, where \\ 
$L_0=6.31$, $\tau=26.3$~days for $0<t<46.0$~days,\\ 
$L_0=1.80$, $\tau=103.7$~days for $46<t<103.5$~days,\\ 
$L_0=13400$, $\tau=140.0$~days for $103.5<t<140.0$~days,\\ 
$L_0=3.51$, $\tau=108.9$~days for $140.0<t<429.5$~days,\\ 
and $L_0=0$ for $t>429.5$~days. $L_0$ is in
units of $10^{42}$~ergs s$^{-1}$. We caution that, owing to the omission of
the SN flux outside the $UBVRI$ range, this input probably
underestimates the true total luminosity.  Shortward of the $U$ band
the input function may be only $\sim$0.5 of the true peak
luminosity, but was probably closer to reality during the plateau
phase.  At wavelengths longer than the $I$~band, the input function may
underestimate the true luminosity by just $\sim$5\% in the peak but up
to 30\% during the plateau.  However, at these wavelengths the
absorptivity of the dust grains (size $\sim$0.1~$\mu$m) is likely to
have fallen well below unity.  In the IR-echo results presented here
we conservatively use just the $UBVRI$ BLC.

Preliminary runs of the IR-echo model showed that, for typical grain
radii (0.05--0.5~$\mu$m), to reproduce the cold component of the SN
SED the dust had to lie at least 10~pc from SN~2004et. In a
simple, spherically symmetric IR-echo model, a dust mass of 
$\sim$350\,M$_{\odot}$ was required to yield the necessary luminosity,
corresponding to a total mass of $\sim$60000\,M$_{\odot}$. Clearly
such a large mass could not have arisen in the progenitor CSM. Indeed,
such a mass would be exceptionally large even for a star-formation
nebula.  A more natural explanation is that the cold component is due
to an IR echo from interstellar, rather than circumstellar or
otherwise local, dust. Supernova-triggered interstellar IR echoes were
first predicted by \citet{be:80} and \citet{wright:80}. 

In our interstellar IR-echo model, we invoked a uniform dust density
extending from 10~pc (see above) to 100~pc from the supernova.
Evaporation by the peak SN luminosity would be unlikely to produce a
cavity exceeding $\sim$0.1~pc.  However, a 10~pc cavity could have
been caused by the progenitor wind or by neighboring stars. The 100~pc
outer limit was chosen as the typical scale height of interstellar
dust in a late-type spiral. (The host galaxy, NGC~6946, is practically
face-on.) For ease of computation, the spherically symmetric geometry
of the IR-echo model was retained. While not appropriate in general
for the outer, presumably planar, limits of the dust in the galactic
disk, for the early era being considered ($\sim$3~yr relative to
the SN-cavity-edge-SN light travel time of 65~yr) spherical
symmetry provides a good approximation; at this
epoch, the echo ellipsoid is extremely elongated (small minor/major
axis ratio), and so the region of the spherical model outer surface
intercepted by the ellipsoid is small and well approximates a plane
surface.  The outer radius is relatively uncritical for the match to
the data: reducing the outer radius by 35\% reduces the 24~$\mu$m
flux by just 20\%.  For ease of computation, we assumed that the grain
material was amorphous carbon where, for wavelengths longer than
$2\pi a$, the grain absorptivity/emissivity can be well approximated
as being proportional to $\lambda^{-1.15}$ \citep{rouleau:91}.  For
shorter wavelengths, an absorptivity/emissivity of unity was used.
Also, for simplicity a single grain size was assumed.  A material
density of 1.85~g~cm$^{-3}$ was adopted \citep{rouleau:91}.

In matching the IR-echo model to the cold residual in the data, the
free parameters were (i) the grain size, which influenced the typical
dust temperatures, and (ii) the grain number density, which determined
the luminosity.  For gas densities of $0.7\pm0.3$~cm$^{-3}$ and
maintaining a 0.006 dust/gas mass ratio, fair fits to the residual
were achieved with grain radii in the range 0.2--0.005~$\mu$m with
corresponding cavity radii of 8--16~pc and optical depths to
UV/optical photons of 0.06--0.1.  For grain radii less than
$\sim$0.02~$\mu$m the absorption cross-section to UV/optical photons
falls below the geometrical value, so for grains smaller than
0.02~$\mu$m the cavity radius had to be fixed at about 16~pc in order
to prevent the dust temperature becoming too low. In addition, for
grain radii less than 0.02~$\mu$m, the total optical depth to
the UV/optical photons converged on a value of about 0.1.  For cavity
radii less than 8~pc, to prevent the dust temperature becoming too
high the grain size had be increased beyond 0.2~$\mu$m with a
consequent increasingly implausible rise in interstellar medium (ISM)
gas density above 1~cm$^{-3}$.  We adopted a grain radius of
0.1~$\mu$m, yielding a grain number density of
$(0.7\pm0.1)\times10^{-12}$~cm$^{-3}$ and an ISM gas density of
$0.6\pm0.1$~cm$^{-3}$.  This is a typical density for the gas disk of
a late-type galaxy such as NGC~6946.  The optical depth to UV/optical
photons is about 0.07, which is easily encompassed within the total
$A_V = 1.0$ mag derived above.  The temperature of the dust grains at
the inner cavity edge is 125~K.

In Fig.\,\ref{fig8}, upper panel, we show the model (silicate IDM)
match for 464~days and note that by 24~$\mu$m the IR echo contributes
more than half the total flux.  In Fig.\,\ref{fig9}, where all of the
total 3-component model matches (silicate IDM) are shown, it is
particularly striking that the long-wave region ($\lambda\geq20~\mu$m)
is well reproduced by an IR-echo model having a {\it single} set of
parameters for all epochs in the period 300--795~days. Even at
days~1125, 1222 and 1395 the long-wave flux is reasonably reproduced
by the IR echo.  We conclude that the cold component is indeed caused
by an IR echo from interstellar dust.

Finally, we note that \citet{wooden:93} included free-free and
free-bound emission in their model of the IR continuum of
SN~1987A. They found that up to about 415~days post-explosion the
free-free radiation dominated the total flux longward of
$\sim$30~$\mu$m. Their model did not include possible emission from an
interstellar IR echo. We examined the possible contribution of
free-free plus free-bound emission to the SN~2004et spectrum.  We
assumed that such emission was dominated by ionised hydrogen in the
SN~2004et ejecta.  The total (ionised+neutral) mass of hydrogen was
assumed to be in the range $1-10$~M$_{\odot}$ and its extent was
inferred from the H$\alpha$ line widths. The number density and
temperature were adjusted to find the maximum free-free plus
free-bound flux which was consistent with the optical to MIR
observations and with the adopted hydrogen mass range.  Within these
constraints, an electron temperature of 4400--4700~K was obtained at
301~days cooling to 3850--4100~K at 795~days. We found that, prior to
the ejecta/CSM collision, free-free emission could account for up to
$\sim$20\% of the total 24~$\mu$m flux at 301~days, declining to less
than 1\% by 796~days.  We conclude that the long wavelength continuum
of SN~2004et was dominated by interstellar IR echo emission.

We would expect the IR-echo luminosity to remain approximately
constant until the vertex of the echo ellipsoid reaches the boundary
of the dust-free cavity.  This would be in a time given by $R_{\rm in} 
= ct/2$, where $R_{\rm in}$ is the cavity radius.  For the 10~pc
dust-free cavity adopted above, and assuming a uniform dust density
10~pc to 100~pc from the supernova, we can expect the echo luminosity
to remain constant for about 65 yr.  We would also expect the
apparent diameter of the echo to grow with time. At the latest epoch
(day~1395) the echo would appear as an annulus having inner and outer
diameters of about $0\farcs3$ and $1\farcs0$, respectively. This is
well below the spatial resolution of {\em Spitzer \/} at
24~$\mu$m. However, for the next $\sim$10~yr the annulus would grow
roughly as $\sqrt t$. Thus, an additional test of the IR-echo
hypothesis would be, in future missions, to attempt to detect and
spatially resolve the MIR emission from SN~2004et.

\section{Conclusion}
We have presented the first-ever comprehensive MIR study of a Type~II-P
supernova, the most common of all SN types. Relative to the fluxes at
optical wavelengths, the MIR luminosity of SN~2004et exhibited a
strong and growing excess between 300 and 1395~days past explosion.
We have shown that this is due to three types of supernova-dust
interactions. We have also presented some of the latest-ever optical
spectra for this type of supernova.

During the period 300-795~days, the SED of SN~2004et is best described
in three parts: (a) a hot component due to emission from optically
thick gas, as well as free-bound radiation, (b) a warm component due to
newly formed, radioactively heated dust in the ejecta, and (c) a cold
component due to an IR echo from the ISM dust of the
host galaxy, NGC~6946. There may also have been a small contribution 
to the IR SED due to free-free emission from ionised gas in the ejecta.
While it is conceivable that the warm component
might alternatively have been caused by an IR echo from dust in the
progenitor CSM, the fine-tuning required to make this scenario work,
plus the blueshift in the [O~I] $\lambda$6300 line after day~300 and the
accelerated decline in the optical light curves, persuades us that a
warm source in newly formed ejecta dust is more likely.

Our modeling of the warm component SED for the era 300--795~days
demonstrates that the dust responsible for the MIR radiation must be
made of silicate material. This is the first time that direct
spectroscopic evidence has been presented for silicate dust formed in
the ejecta of a recent supernova.  Additional support for the silicate
scenario comes from our detection of a large, but declining, mass of
SiO, a key compound in the silicate-formation sequence. The mass of
directly detected dust grew to no more than a few times
$10^{-4}\,{\rm M}_{\odot}$.

A remarkable result of our analysis is that the model radius remained
constant (standard deviation $<10$\%) throughout the 300--795~day era, 
yet the SN itself expanded by a factor of 2.65.  We propose that 
this constitutes evidence
that the dust formed in small, dense, comoving clumps, similar to the
scenario argued for SN~1987A.  Such clumps might arise in localized
regions of density enhancement.  With more efficient cooling in such
regions, the clumps might be unable to expand against the hotter,
higher pressure interclump gas.  We investigated the case of a flat
distribution of uniform-density clumps and showed that such a scenario
could indeed account for the apparent constancy of the radius of the
IR emission region.  Moreover, we found that the derived dust masses
are much the same as in the case of a uniform cloud of dust; the
invocation of uniform-density clumps does not allow us to ``hide''
larger masses of dust. 

We have shown that the cold residual component is well reproduced by a
single IR-echo model, where the SN UV/optical flux is reradiated by
the interstellar dust of NGC~6946. The model match plus the demand
that the ISM number density should not exceed 
1~cm$^{-3}$ (assuming a dust/gas mass ratio of $\sim$0.006) implies 
a dust-free cavity of radius 8--16~pc, possibly caused by the 
stellar winds of the progenitor and other nearby stars. While we 
assumed that the interstellar grains
are made of amorphous carbon, we do not rule out other grain
materials.  The important point about our ISM IR~echo model is that
with a single pair of input parameters (the dust-free cavity
radius and the ISM dust density), we can account for the cold component
throughout the 300--1395~day era. This gives us confidence that the ISM
IR-echo interpretation is correct.

An interesting prediction from this result is that the echo luminosity
should remain high for some decades. Moreover, the diameter of the
echo annulus on the sky should increase to several arcseconds over the
next $\sim$10 yr, potentially resolvable by upcoming MIR space
missions.  We note also that the occurrence of such cold IR echoes
should be relatively common for SNe occurring in dusty, late-type
galaxies. \citet{meikle:07} have already pointed out that such an echo
provides a natural explanation for the cool MIR flux at days~670--681
from SN~2003gd.

After $\sim$1000~days, we observed a remarkable rise in the optical
and MIR fluxes. In addition, by day~823 the optical spectral lines
had developed wide, box-shaped profiles with evidence of dust
attenuation within the line-producing region. We interpret these
results as being due to the impact of the ejecta on the progenitor
CSM, with the MIR flux rise and optical line attenuation resulting
from dust formation within the cool dense shell. We estimate a 
CDS dust mass of $(2-5)\times10^{-4}\,{\rm M}_{\odot}$, a CDS 
radius of $\sim$6000~AU, and a total mass of 
$0.03-0.08\,{\rm M}_{\odot}$. For a red supergiant wind this
mass is entirely plausible. Also, for a typical velocity of 10\,\kms\,
the radius indicates a CSM age of $\sim$2700~yr.

The work presented here adds to the growing number of studies
which do not support the contention that SNe are responsible for the
large masses of dust in high-redshift galaxies.  There may still be some 
possibility for the existence of large masses of undetected ejecta dust in 
young SNe. Such dust might (a) be formed after the end of the observations,
(b) be hidden in nonuniform density distributions, or (c) exist at
very low temperatures and so go undetected in MIR/FIR studies. But
there are difficulties with each scenario. Option (a) would be
in conflict with dust-condensation models \citep{todini:01,nozawa:03}
which suggest that the bulk of the dust condensation is complete
within two years past explosion.  Our study, which includes the
effects of clumping, reduces the plausibility of option (b). Possibility 
(c) requires most of the dust formed in the ejecta to rapidly cool to 
less than a few 100~K. It is not clear if this is physically possible, 
nor whether such a scenario could be made to be consistent with the
optical depths indicated by the MIR spectra.

This study has demonstrated the rich, multi-faceted ways in which a
typical core-collapse supernova and its progenitor can produce and/or
interact with cosmic dust. The range of distance scales over which we
observe these processes is also noteworthy: typically 300~AU for the
ejecta dust, 6000~AU for the CDS, and over $2\times10^6$~AU for the IS
IR~echo.  We have shown that, at least in the case of silicate dust,
through the use of MIR spectra we can estimate total masses of dust
even in uniform-density clumps which are optically thick 
{\it in the MIR}.  We find that the mass of directly observed dust 
produced either in the ejecta or CDS never exceeded $10^{-3}\,{\rm
 M}_{\odot}$. This is consistent with a steadily increasing number of
similar results.

\acknowledgements 
We thank R. M. Crockett and collaborators for their permission to use
their optical observations of the SN~2004et field prior to
publication.  This work is based on observations made with the Spitzer
Space Telescope, which is operated by the Jet Propulsion Laboratory,
California Institute of Technology under a contract with NASA. Support
for this work was provided by NASA through an award issued by
JPL/Caltech. A.V.F. gratefully acknowledges additional support from
NSF grant AST--0607485. J.S. is a Royal Swedish Academy of Sciences
Research Fellow supported by a grant from the Knut and Alice
Wallenberg Foundation. S.M. acknowledges support from the Academy of
Finland (project 8120503). The Dark Cosmology Centre is funded by the
Danish National Research Foundation.  We thank the SUSPECT team for
continued maintenance of their excellent database.  Some of the data
presented herein were obtained at the W. M. Keck Observatory, which is
operated as a scientific partnership among the California Institute of
Technology, the University of California, and NASA; it was made
possible by the generous financial support of the W. M. Keck
Foundation.  We wish to extend special gratitude to those of Hawaiian
ancestry on whose sacred mountain we are privileged to be guests.
A.V.F. thanks the Aspen Center for Physics, where he participated in a
workshop on Wide-Fast-Deep Surveys while this paper was nearing
completion.

~\\
~\\
~\\

\newpage

\begin{table*}
\caption{Background-Subtracted Mid-IR Photometry of SN~2004et
\label{tab:phot}}
\begin{center}
\begin{tabular}{lccclcccccc}
\hline
        &  & & & \multicolumn{7}{c}{Flux ($\mu$Jy)} \\
\cline{5-11}
& & Epoch$^\dag$ & t$_{\mathrm{exp}}$ & \multicolumn{4}{c}{IRAC} & \multicolumn{2}{c}{PUI} & MIPS \\
Date    & MJD & (d)    & (s) & 3.6\,$\mu$m  & 4.5$\mu$m & 5.8\,$\mu$m  & 8.0\,$\mu$m &  16\,$\mu$m & 22\,$\mu$m & 24\,$\mu$m       \\
\hline
2004 Jun. 10$^a$  & 53166.8 &--103.2& 107    & 84.9(4.1)&47.5(4.9)&165(17)   &351(45)  & \nodata & \nodata &\nodata   \\
2004 Jul. 09$^a$  & 53196.0 &--74.0 & 81     & \nodata  &\nodata  &\nodata   &\nodata  & \nodata & \nodata &  268(46)     \\
2004 Jul. 11$^a$  & 53197.2 &--72.8 & 81     & \nodata  &\nodata  &\nodata   &\nodata  & \nodata & \nodata &  207(37) \\\hline
2004 Nov. 25$^a$  & 53334.7 & 64.7  &107&17020(30)&13200(20)&  9490(30)& 5950(30)& \nodata & \nodata & \nodata  \\
2005 Jul. 13$^b$  & 53564.9 & 294.9 & 629    & \nodata  & \nodata & \nodata  & \nodata & 676(49) & \nodata & \nodata  \\
2005 Jul. 19$^c$  & 53570.9 & 300.9 & 14     &   686(17)& 2994(20)& 1344(50) & 2277(40)& \nodata & \nodata & \nodata  \\
2005 Jul. 20$^b$  & 53571.2 & 301.2 & 536    &   697(3) &  3040(5)& 1446(13) & 2252(29)& \nodata & \nodata & \nodata  \\
2005 Aug. 03$^c$  & 53585.5 & 315.5 & 159    & \nodata  & \nodata & \nodata  & \nodata & \nodata & \nodata & 670(50)  \\
2005 Sep. 17$^c$  & 53630.8 & 360.8 & 14 &   475(12)& 1725(20)& 1064(48) & 1930(55)& \nodata & \nodata & \nodata  \\
2005 Sep. 24$^c$  & 53637.0 & 367.0 & 159    & \nodata  & \nodata & \nodata  & \nodata & \nodata & \nodata & 685(45)  \\
2005 Nov. 02$^b$  & 53676.0 & 406.0 & 536    &   319(3) &  1148(3)&  706(13) & 1600(25)& \nodata & \nodata & \nodata  \\
2005 Dec. 22$^b$  & 53726.5 & 456.5 & 629    & \nodata  & \nodata & \nodata  & \nodata & 657(45)& \nodata & \nodata   \\
2005 Dec. 30$^c$  & 53734.9 & 464.9 & 14 &   193(15)&  640(19)&  540(70) & 1224(43)& \nodata & \nodata & \nodata  \\
2006 Jan. 10$^c$  & 53745.7 & 475.7 & 159    & \nodata  & \nodata & \nodata  & \nodata & \nodata & \nodata & 667(65)  \\
2006 Aug. 04$^d$  & 53951.3 & 681.3 & 315    & \nodata  & \nodata & \nodata  & \nodata &  434(48) & 791(50)& \nodata  \\
2006 Aug. 13$^e$  & 53960.5 & 690.5 & 29     & $<35$ (2$\sigma$)&33(15)&150(60)& 430(30)& \nodata& \nodata & \nodata  \\
2006 Sep. 01$^e$  & 53979.2 & 709.2 & 159    & \nodata  & \nodata & \nodata  & \nodata & \nodata & \nodata & 480(65)  \\
2006 Sep. 10$^e$  & 53988.9 & 718.9 & 57     & \nodata  & \nodata &\nodata   & \nodata & 296(47) & \nodata & \nodata  \\
2006 Sep. 28$^d$  & 54006.2 & 736.2 & 268    &  15(5)   & 52(3)   & $<67$ (2$\sigma$)& 248(18)& \nodata & \nodata & \nodata \\
2006 Oct. 18$^d$  & 54026.5 & 756.5 & 629    & \nodata  & \nodata & \nodata  & \nodata & 416(46) & 742(55) & \nodata \\
2006 Nov. 26$^d$  & 54065.9 & 795.9 & 536    & $<5$ (2$\sigma$) & 22(3)& $<40$ (2$\sigma$) & 172(19)  & \nodata & \nodata & \nodata \\
2006 Dec. 29$^e$  & 54098.0 & 828.0 & 250    & $<5$ (2$\sigma$) & 38(3)& $<35$ (2$\sigma$) & 65(32)*   & \nodata & \nodata & \nodata \\
2007 Jan. 21$^e$  & 54121.2 & 851.2 & 494    & \nodata  & \nodata & \nodata  & \nodata & \nodata & \nodata & 600(65)     \\
2007 Jan. 27$^e$  & 54127.5 & 857.5 & 132    & \nodata  & \nodata &\nodata   & \nodata &   251(48)   & \nodata & \nodata \\
2007 Jun. 26$^e$  & 54277.1 &1007.1 & 283    & \nodata  & \nodata &\nodata   & \nodata &  100(48)   & \nodata & \nodata  \\
2007 Jul. 04$^e$  & 54285.0 &1015.0 & 283    & $<3$ (2$\sigma$)& 20.7(2.8) &$<30$ (2$\sigma$) & 109(19)& \nodata & \nodata & \nodata \\
2007 Jul. 10$^e$  & 54291.1 &1021.1 & 494    & \nodata  & \nodata & \nodata  & \nodata & \nodata & \nodata & 511(59)     \\
2007 Aug. 02$^f$  & 54314.3 &1044.3 & 283    & \nodata  & \nodata &\nodata   & \nodata &  149(45)  & \nodata & \nodata   \\
2007 Aug. 12$^f$  & 54324.4 &1054.4 & 322    &  $<4$ (2$\sigma$)&  16.2(2.5) & $<30$ (2$\sigma$) & 134(35) & \nodata & \nodata & \nodata \\
2007 Aug. 27$^g$  & 54339.7 &1069.7 &  93    & \nodata  & \nodata & \nodata  & \nodata & \nodata & \nodata & 554(30)     \\
2007 Aug. 27$^f$  & 54339.7 &1069.7 & 494    & \nodata  & \nodata & \nodata  & \nodata & \nodata & \nodata & 480(35)     \\
2007 Oct. 12$^g$  & 54385.2 &1115.2 &1258    & \nodata  & \nodata & \nodata  & \nodata & 956(54)& \nodata & \nodata      \\
2007 Oct. 22$^g$  & 54395.9 &1125.9 &3485    & 66.4(1.5)& 189.9(2.6)&335(9) & 373(31) & \nodata & \nodata & \nodata      \\
2007 Dec. 27$^f$  & 54461.1 &1191.1 & 322    & 52.2(2.1)& 197.0(2.8)& 320(12) & 444(37)& \nodata & \nodata & \nodata   \\
2008 Jan. 07$^f$  & 54472.5 &1202.5 & 494    & \nodata  & \nodata & \nodata  & \nodata & \nodata & \nodata & 1192(74)             \\
2008 Jan. 17$^g$  & 54482.2 &1212.2 &1258    & \nodata  & \nodata & \nodata  & \nodata & 962(45) & \nodata & \nodata      \\
2008 Jan. 27$^g$  & 54492.8 &1222.8 & 536    & 53.8(1.2)&  219(3) & 360(14)  & 335(25) & \nodata & \nodata & \nodata     \\
2008 Feb. 14$^g$  & 54510.6 &1240.6 &  93    & \nodata  & \nodata & \nodata  & \nodata & \nodata & \nodata &  1195(55)  \\
2008 Jul. 18$^f$  & 54665.8 &1395.8 & 322    & 85.7(1.5)&  270(2) & 456(8)   & 633(23) & \nodata & \nodata & \nodata    \\
2008 Jul. 29$^f$  & 54676.0 &1406.0 & 494    & \nodata  & \nodata & \nodata  & \nodata & \nodata & \nodata & 1461(26) \\
\hline
Background                     & \nodata&\nodata&\nodata   & 84.9(4.1)&47.5(4.9)&165(17)   &351(45)  & 345(40)**  & 250(40)** & 238(30) \\
\hline
\end{tabular}
\end{center}
\tablecomments{
Statistical uncertainties in the last one or two significant figures are 
shown in brackets.\\
$^\dag$ Following \citet{li:05}, we assume an explosion date of 2004 
Sep. 22.0 (MJD = 53270.0).\\
$^a$PID. 00159 Kennicutt et al. (SINGS).\\
$^b$PID. 20256 Meikle et al. (MISC).\\
$^c$PID. 20320 Sugerman et al. (SEEDS).\\
$^d$PID. 30292 Meikle et al. (MISC).\\
$^e$PID. 30494 Sugerman et al. (SEEDS).\\
$^f$PID. 40010 Meixner et al. (SEEDS).\\
$^g$PID. 40619 Kotak et al. (MISC).\\
$^*$Owing to the weakness of the $8~\mu$m flux at this epoch and the 
consequent difficulty of correcting for the strong, complex background 
via the usual annular sky measurement, the background was, instead, 
measured using five $5''$-radius apertures around the target 
aperture.  This allowed the exclusion of the worst irregularity in 
the residual background. \\
$^{**}$ No pre-explosion PUI data available. Background estimated by 
blackbody interpolation between pre-explosion IRAC and MIPS photometry. 
\label{tab:phot}}
\end{table*}

\begin{table}
\caption{Mid-IR Spectroscopy of SN~2004et} 
\begin{center}
\begin{tabular}{lcccccc}
\hline
 Date            & MJD     & Epoch$^\dag$ & \multicolumn{4}{c}{Spectral range} \\ 
 (UT)            &         &  (d)  &   5.2  &   7.4 &   14.0 & 19.6 \\ 
                 &         &       & --8.7  &--14.5 & --21.3 & --35       \\ 
                 &         &       & $\mu$m    &   $\mu$m  &  $\mu$m  &  $\mu$m \\ \hline
 2005 Jul. 13$^a$ &53564.87 & 294.9 &    x            &    x             &     x             &     x \\      
 2005 Sep. 06$^b$ &53619.77 & 349.8 &    x            &    x             &     x             &     x \\ 
 2005 Dec. 16$^b$ &53720.77 & 450.8 &    x            &    x             &     x             &    -- \\ 
 2006 Jan. 16$^a$ &53751.01 & 481.0 &    x            &    x             &     x$^{\dag\dag}$  &     x$^{\dag\dag}$ \\ 
 2006 Dec. 24$^c$ &54093.12 & 823.1 &    x            &    x             &    --             &    -- \\ 
 2008 Jan. 17$^d$ &54482.20 &1212.2 &    x            &    x             &     x             &     x \\  
 2008 Jul. 08$^e$ &54655.60 &1385.6 &    x            &    x             &     x             &     x \\    
\hline
\end{tabular}
\end{center}
\tablecomments{Spectral ranges observed at each epoch are indicated 
by ``x.'' These correspond to the Short-Low (SL) first and second 
orders (7.4--14.5 and 5.2--8.7\,$\mu$m, respectively) and the 
Long-Low (LL) first and second orders (19.6--38 and 14.0--21.3\,$\mu$m, 
respectively). \label{tab:speclog}\\
$^\dag$ Following \citet{li:05}, we assume an explosion date of 
2004 Sep. 22.0 (MJD = 53270.0).\\
$^{\dag\dag}$ The LL spectra from day~481 were of low signal-to-noise 
ratio and contaminated by residual background. They are not shown in 
Fig. \ref{fig3} nor used in the analysis. \\
$^a$PID. 20320 Sugerman et al. (SEEDS).
$^b$PID. 20256 Meikle et al. (MISC).
$^c$PID. 30292 Meikle et al. (MISC).
$^d$PID. 40619 Kotak et al. (MISC).
$^e$PID. 40010 Meixner et al. (SEEDS).
\label{speclog}}
\end{table}

\begin{table*}
\caption{Late-Time Optical Magnitudes of SN~2004et}
\begin{center}
\begin{tabular}{lccccccl}
\hline
              &       & Epoch$^\dag$ & \multicolumn{4}{c}{Magnitudes}& Source \\
Date          & MJD   & (d)    & $B$ & $V$ & $R$ & $I$ & \\
\hline
2006 Dec. 24  & 54093 & 823    & 22.65(20)  & 22.15(20) & 21.65(20) & 21.15(20) & Keck spectrum \\
2007 Apr. 13  & 54203 & 933    &\nodata$^a$ & 22.25(20) & 21.55(20) & 21.20(20) & Keck spectrum  \\
2007 Jul. 08  & 54289 &1019    & 23.40(10)  & 23.20(10) & 23.05(10) & 22.85(10) & {\it HST} \\
2007 Aug. 12  & 54324 &1054    & 24.00(10)  & 23.49(8) & 22.59(8) & 22.55(12) & WHT \\ 
2007 Nov. 12  & 54416 &1146    & 22.85(20)  & 22.55(20) & 22.00(20) & 21.70(20) &  Keck spectrum \\
\hline
Nearby stars  &       &        & 25.00(10)  & 24.30(10) & 23.75(10) & 23.00(10)  & {\it HST}   \\
\hline
\end{tabular}
\end{center}
\tablecomments{Magnitudes derived from spectra obtained
at the Keck telescope and from \citet{crockett:09} ({\it HST} and 
WHT). Uncertainties in the last two figures are shown in
brackets.  The SN magnitudes have been corrected for the effects of
two faint stars which lay within the seeing disk of the ground-based
observations. The bottom line shows the combined magnitudes of the two
stars derived from the {\it HST} magnitudes reported by \citet{crockett:09}
(see text).\\
$^\dag$ Following \citet{li:05}, we assume an explosion date
of 2004 Sep. 22.0 (MJD = 53270.0).\\ 
$^a$There is no $B$ value on day~933 since at this epoch there was
only partial coverage of the appropriate section of the spectrum.
\label{tab:keckphot}}
\end{table*}

\begin{table}
\caption{Blackbody Parameters for Matches to SN 2004et SEDs} 
\begin{center}
\begin{tabular}{cccccccccccc}
\hline
Epoch &$v_{\mathrm{hot}}$ &$T_{\mathrm{hot}}$& $v_{\mathrm{warm}}$&$R_{\mathrm{warm}}$&$T_{\mathrm{warm}}$&$v_{\mathrm{cold}}$&$T_{\mathrm{cold}}$&$L_{\mathrm{rad}}$&  $L_{\mathrm{hot}}$&$L_{\mathrm{warm}}$&  $L_{\mathrm{cold}}$ \\
  (d) &   (\kms) &    (K)  &    (\kms) &($10^{16}$cm)&(K)  &   (\kms) &   (K)   & ($10^{38}$  &($10^{38}$&($10^{38}$& ($10^{38}$\\
      &          &         &           &             &     &          &         &\ergs)       & \ergs)   & \ergs)   &\ergs)     \\\hline
  64  &   3000   &   5300  &  \nodata  & \nodata  &\nodata & \nodata   &\nodata & 4030 &    15470     &  \nodata     &   \nodata\\
 300  &     66   &   6000  &    1600   &0.41   &    700    &    12000  &   130  &  485 &      270     &       29     &     1.3 	\\
 360  &     38   &   6500  &    1300   &0.40   &    670    &    12000  &   130  &  255 &      180     &       23     &     1.4 	\\      
 406  &     28   &   6300  &    1250   &0.44   &    610    &    11000  &   130  &  160 &      110     &       19     &     2.0 	\\ 
 464  &     14   &   7000  &    1100   &0.44   &    570    &     9500  &   130  &   86 &       54     &       15     &     2.6 	\\
 690  &     3.3  &   6500  &     850   &0.51   &    410    &     9000  &   110  &   7.6  &    4.9     &      5.2     &     2.4 	\\
 736  &     2.6  &   7000  &     830   &0.53   &    400    &     8500  &   110  &   4.8  &    4.7     &      5.1     &     2.7 	\\
 795  &     2.2  &   7000  &     650   &0.45   &    390    &     6500  &   120  &   2.6  &    3.9     &      3.3     &     2.9 	\\
1125  &     0.62 &  10000  &     510   &0.50   &    500    &     6000  &   120  &   0.17 &    2.6     &     10.9     &     5.0 	\\
1222  &     0.83*&  10000* &     500   &0.53   &    500    &     7000  &   120  &   0.10 &    5.5*    &     12.4     &     8.1  \\
1395  &     1.07*&  15000* &     465   &0.56   &    510    &     6000  &   120  &   0.05 &   60.0*    &     15.1     &     7.7  \\\hline      
\end{tabular}
\tablecomments{The blackbody luminosities ($L_{\mathrm{hot}}$,
$L_{\mathrm{warm}}$, $L_{\mathrm{cold}}$) are given in col. 10--12.
In col.~9, $L_{\mathrm {rad}}$ is the radioactive deposition
luminosity corresponding to the ejection of 0.055\,M$_{\odot}$ of
$^{56}$Ni, scaled from the SN~1987A case described by \citet{li:93}.\\
$^*$The day~1222 and day~1395 optical parameters are approximate. They
were obtained by blackbody matches to fluxes obtained through
extrapolation of the optical light curves.}
\label{tab:bb}
\end{center}
\end{table}

\begin{table*}
\caption{Model (Silicate) Parameters for Matches to SN 2004et Observations} 
\begin{center}
\begin{tabular}{cccccccccccc}
\hline
Epoch &$v_{\mathrm{hot}}$&$T_{\mathrm{hot}}$ &$v_{\mathrm{warm}}$& $R_{\mathrm{warm}}$&$T_{\mathrm{warm}}$&$\tau_{10\mu m}$&$L_{\mathrm{rad}}^\dag$&$L_{\mathrm{hw}}^{\dag\dag}$& $L_{\mathrm{warm}}$ & $M_{\mathrm{dust}}$  &$M_{\mathrm{SiO}}$\\
  (d) &   (\kms) &    (K)  &    (\kms)&($10^{16}$\,cm) &   (K)   &  &($10^{38}$&($10^{38}$&($10^{38}$&($10^{-4}\,{\mathrm{M}}_{\odot}$)&($10^{-4}\,{\mathrm{M}}_{\odot}$)\\ 
      &          &         &          &                &         &  &  \ergs)  &  \ergs)  & \ergs)   &                                 &                                 \\\hline    
 300  &     61   &   6000  &    1700  & 0.44 &    900   &      2.8       &   485       &   271   & 40.5 &     0.39   &   5.7      \\
 360  &     38   &   6500  &    1600  & 0.50 &    730   &      3.1       &   255       &   202   & 25.4 &     0.57   &   5.8      \\      
 406  &     25   &   6700  &    1420  & 0.50 &    700   &      3.3       &   160       &   132   & 22.0 &     0.60   &   4.1 \\
 464  &     14   &   7000  &    1250  & 0.50 &    650   &      3.6       &    86       &   71.6  & 17.7 &     0.66  &   3.7      \\
 690  &     3.3  &   6500  &     800  & 0.48 &    500   &      5.2       &   7.6       &   12.5  &  7.1 &     0.85  &$\sim$0.9   \\
 736  &     2.6  &   7000  &     680  & 0.43 &    450   &      6.6       &   4.8       &    9.5  &  4.3 &     0.90  & $\sim$0.6  \\
 795  &     2.2  &   7000  &     620  & 0.43 &    400   &     11.5       &   2.6       &    6.5  &  3.3 &     1.5   &$\sim$0.5 \\
\hline
\end{tabular}
\label{tab:sio}
\end{center}
\tablecomments{
$^\dag$ $L_{\mathrm{rad}}$ is the radioactive deposition luminosity.\\
$^{\dag\dag}$ $L_{\mathrm{hw}} = L_{\mathrm{hot}} + L_{\mathrm{warm}}$\\
As discussed in the text, by day~1125 our model is no longer
appropriate; hence, we do not list the parameters for days~1125, 1222 
or 1395 above. Typical silicate model uncertainties:\\
day~300: $v_{\mathrm{warm}}=1700^{+300}_{-200}$~km/s, $T_{\mathrm{warm}}=900^{+100}_{-140}$~K, $M_{\mathrm{dust}}=(0.39^{+0.25}_{-0.12})\times10^{-4}$~M$_{\odot}$\\
day~464: $v_{\mathrm{warm}}=1250^{+150}_{-100}$~km/s, $T_{\mathrm{warm}}=650^{+50}_{-50}$~K, $M_{\mathrm{dust}}=(0.66^{+0.27}_{-0.15})\times10^{-4}$~M$_{\odot}$\\
day~795: $v_{\mathrm{warm}}=620^{+130}_{-70}$~km/s, $T_{\mathrm{warm}}=400^{+20}_{-40}$~K, $M_{\mathrm{dust}}=(1.5^{+1.2}_{-0.4})\times10^{-4}$~M$_{\odot}$}
\end{table*}

\begin{table*}
\caption{Parameters for Cloud of Silicate Dust Clumps Model Matches}
\begin{center}
\begin{tabular}{cccccccc}
\hline
Epoch &$T_{\mathrm{warm}}$&$R_{\mathrm{cl(max)}}$&$f$&$\tau_{\mathrm{cl(max)}}$&$\tau_{\mathrm{cd}}$& M$_{\mathrm{dust}}$  &M$_{\mathrm{SiO}}$\\
  (d) &  (K)  &($10^{14}$cm)&($\times100$)&  at $10~\mu$m    & at $10~\mu$m & ($10^{-4}\,{\rm M}_{\odot}$)&($10^{-4}\,{\rm M}_{\odot}$)\\ \hline
 300  &  900  & 1.83  &  1.4     &   2.6  & 0.43  &  0.40  &   5.7    \\
 360  &  730  & 1.95  &  1.0     &   3.5  & 0.36  &  0.61  &   5.8    \\      
 406  &  700  & 2.10  &  1.84     &   3.0  & 0.32  &  0.61  &   4.1    \\
 464  &  650  & 1.91  &  0.43    &   4.1  & 0.21  &  0.69  &   3.7    \\
 690  &  500  & 1.63  &  0.084    &   8.2  & 0.07  &  0.99  &$\sim$0.9 \\
 736  &  450  & 1.43  &  0.046   &  11.6  & 0.05  &  1.1  & $\sim$0.6\\
 795  &  400  & 1.53  &  0.044   &  13.0  & 0.05  &  1.4   &$\sim$0.5 \\
\hline
\end{tabular}
\label{tab:clo}
\end{center}
\tablecomments{With $m_1=0$, $R_{\rm cl(min)} = 0.01R_{\rm cl(max)}$, 
$v_{\rm warm} = 2500$ \kms, and $N_{\rm cl}=2500$.}
\end{table*}


\vspace*{2cm}
\appendix

\section{Infrared Radiation from a Cloud of Dust Clumps}
Consider a spherical ``cloud'' of uniform clumps, with a cloud radius
of $R_{\rm cd}$. A clump-size distribution $dn_{\rm cl} = 
k_1R_{\rm cl}^{-m_1}dR_{\rm cl}$ is invoked, where $dn_{\rm cl}$ is the
number density of clumps having radius $R_{\rm cl} \to 
R_{\rm cl}+dR_{\rm cl}$ and $k_1$ is the clump number density 
scaling factor (i.e., it controls the
total number of clumps in the cloud). At any given time, each clump has
the same uniform number density of dust grains.\\

In the original IDM \citep{meikle:07}, we can re-express the
luminosity equation as \begin{equation} L_{cl}(\nu)= 4\pi B(\nu,
T)\sigma_{cl}(\nu),\\ \end{equation} \noindent where the effective
cross section of the dust sphere, $\sigma_{\rm cl}(\nu)$, is given by
\begin{displaymath} \sigma_{\rm cl}(\nu)=\pi
R_{\rm cl}^2(1-e^{-\tau'_{\rm cl}(\nu)})\\ =\pi
R_{\rm cl}^2[0.5\tau_{\rm cl}(\nu)^{-2}(2\tau_{\rm cl}(\nu)^2-1+(2\tau_{\rm cl}(\nu)+1)e^{-2\tau_{\rm cl}(\nu)}],\\
\end{displaymath} \noindent where $\tau'_{\rm cl}(\nu)$ is the effective
optical depth of the dust sphere at frequency $\nu$ and
$\tau_{\rm cl}(\nu)$ is the surface to center optical depth of the sphere
at frequency $\nu$. Now $\tau_{\rm cl}(\nu) = \gamma R_{\rm cl}$, 
where $\gamma=\frac{4}{3}\pi
k\rho\kappa(\nu)\frac{1}{4-m}[a^{4-m}_{\rm (max)}-a^{4-m}_{\rm (min)}]$
and where $\rho$ and $\kappa(\nu)$ are, respectively, the density and mass
absorption coefficient of the grain material \citep{meikle:07}.\\

The luminosity of a single grain, $L_g(\nu)$,
is given by  $L_g(\nu)=4\pi^2 a^2Q(\nu)B(\nu,T)$. Using this to eliminate
$4\pi B(\nu,T)$ from equation A1, we obtain
\begin{displaymath}
L_{\rm cl}(\nu)=L_g(\nu)\sigma_{\rm cl}(\nu)/(\pi a^2Q(\nu))=L_g(\nu)(\sigma_{\rm cl}(\nu)/\sigma_g(\nu)),
\end{displaymath}
where $\sigma_g(\nu)=\pi a^2Q(\nu)$ is the grain absorption cross section.
By analogy, we can write for the cloud of clumps that
\begin{equation}
L_{\rm cd}(\nu)=L_{\rm cl}(\nu)(\sigma_{\rm cd}(\nu)/\sigma_{\rm cl}(\nu))=L_g(\nu)(\sigma_{\rm cd}(\nu)/\sigma_g(\nu)),\\
\end{equation}
where
\begin{equation}
\sigma_{\rm cd}(\nu)=\pi R_{\rm cd}^2(1-e^{-\tau'_{\rm cd}(\nu)})\\
=\pi R_{\rm cd}^2[0.5\tau_{\rm cd}(\nu)^{-2}(2\tau_{\rm cd}(\nu)^2-1+(2\tau_{\rm cd}(\nu)+1)e^{-2\tau_{\rm cd}(\nu)}],\\
\end{equation}
$\tau'_{\rm cd}(\nu)$ is the effective optical depth of the cloud at frequency
$\nu$, and $\tau_{\rm cd}(\nu)$ is the surface to center optical depth
of the cloud at frequency $\nu$.\\

To determine $\tau_{\rm cd}(\nu)$, let us consider the case where the
cloud contains a number of uniform-density clumps with a size
distribution, $dn_{\rm cl}=k_1R_{\rm cl}^{-m_1}dR_{\rm cl}$, as given above.
Thus

\begin{displaymath}
d\tau_{\rm cd}(\nu)= R_{\rm cd}\pi R_{\rm cl}^2(1-e^{-\tau'_{\rm cl}(\nu)})k_1R_{\rm cl}^{-m_1}dR_{\rm cl}.\\
\end{displaymath}

Integrating from $R_{\rm cl(min)}$ to $R_{\rm cl(max)}$, we have\\
\begin{displaymath}
\tau_{\rm cd}(\nu)
  = \pi R_{\rm cd}k_1 \int ^{R_{\rm cl(max)}} _{R_{\rm cl(min)}} (1-e^{-\tau'_{\rm cl}(\nu)})R_{\rm cl}^{2-m_1}dR_{\rm cl}. \\
\end{displaymath}
\begin{displaymath}
  = \pi R_{\rm cd}k_1 \int ^{R_{\rm cl(max)}} _{R_{\rm cl(min)}} [0.5\tau_{\rm cl}(\nu)^{-2}(2\tau_{\rm cl}(\nu)^2-1+(2\tau_{\rm cl}(\nu)+1)e^{-2\tau_{\rm cl}(\nu)}]R_{\rm cl}^{2-m_1}dR_{\rm cl}\\
\end{displaymath}
\begin{displaymath}
  = \pi R_{\rm cd}k_1 \int ^{R_{\rm cl(max)}} _{R_{\rm cl(min)}} \frac{1}{2\gamma^2}(2\gamma^2 R_{\rm cl}^2-1+(1+2\gamma R_{\rm cl})e^{-2\gamma R_{\rm cl}})r_{\rm cl}^{-m_1}dR_{cl}.
\end{displaymath} \\

For the case where $m_1=0$ (i.e., a flat clump-size distribution), this integrates to

\begin{displaymath}
\tau_{\rm cd}(\nu)=\frac{1}{2\gamma^3}\pi R_{\rm cd}k_1[(\frac{2}{3}\gamma^3 R_{\rm cl}^3 - \gamma(1+e^{-2\gamma R_{\rm cl}})R_{\rm cl}-e^{-2\gamma R_{\rm cl}})]_{R_{\rm cl(min)}}^{R_{\rm cl(max)}}.
\end{displaymath}
Substituting into equation A3, we obtain $\sigma_{\rm cd}(\nu)$. This can 
then be used to obtain
$L_{\rm cd}(\nu)$ via equation A2. It may also be shown that the 
total mass of dust in the cloud when $m_1=0$ is given by
\begin{displaymath}
M_{\rm cd}=\frac{16}{27} \pi^3 R_{\rm cd}^3 \rho kk_1 \frac{1}{4-m}(a_{\rm (max)}^{4-m}-a_{\rm (min)}^{4-m})(R_{\rm cl(max)}^4 - R_{\rm cl(min)}^4),
\end{displaymath}
and the total number of dust clumps is
\begin{displaymath}
N_{\rm cl}=\frac{4\pi}{3(1-m_1)}R_{\rm cd}^3k_1(R_{\rm cl(max)} - R_{\rm cl(min)}).
\end{displaymath} 

\end{document}